\documentclass[twocolumn, trackchanges, 12pt]{aastex63}

\usepackage{graphicx,latexsym,amssymb,epsfig}
\usepackage{multirow,amsmath,array,booktabs}
\usepackage{natbib}
\usepackage{url}
\usepackage{subfigure}
\usepackage{tabularx}
\usepackage{color}
\usepackage{bm}
\usepackage{color}
\usepackage{epstopdf}
\usepackage[figuresright]{rotating}

\def\fm3{\;\textrm{fm}^{-3}}

\begin{document}

\title{Influence of effective nucleon mass on equation of state for supernova simulations and neutron stars}

\author{Shuying Li}
\affiliation{School of Physics, Nankai University, Tianjin 300071, People's Republic of China}

\author{Junbo Pang}
\affiliation{School of Physics, Nankai University, Tianjin 300071, People's Republic of China}

\author[0000-0003-2717-9939]{Hong Shen}
\affiliation{School of Physics, Nankai University, Tianjin 300071, People's Republic of China}
\email{shennankai@gmail.com}

\author[0000-0002-1709-0159]{Jinniu Hu}
\affiliation{School of Physics, Nankai University, Tianjin 300071, People's Republic of China}
\email{hujinniu@nankai.edu.cn}

\author[0000-0002-9224-9449]{Kohsuke Sumiyoshi}
\affiliation{National Institute of Technology, Numazu College, Shizuoka 410-8501, Japan}

\begin{abstract}
We investigate the influence of the effective nucleon mass on the equation of state (EOS),
which is constructed for simulations of core-collapse supernovae and
binary neutron star mergers, within the relativistic mean-field (RMF) framework.
The study introduces a new RMF parameter set, TM1m,
which is a modification of the TM1e model with an adjusted effective mass,
maintaining the saturation properties of nuclear matter.
The TM1m model, with a larger effective mass ratio ($M^{\ast}/M \sim 0.8$)
compared to the TM1e model ($M^{\ast}/M \sim 0.63$), is employed to construct
a new EOS table, EOS5. This EOS table is designed to offer insights into
the influence of the effective nucleon mass on the EOS within a relativistic
framework, particularly above the saturation density. The results of EOS5 are compared with those obtained from other models,
including both relativistic and nonrelativistic approaches.
The properties of cold neutron stars, calculated using the TM1m model,
are compatible with the existence of a $2\ M_\odot$ pulsar
and the latest constraints on the tidal deformability and radii
of a canonical $1.4\ M_\odot$ neutron star,
derived from astrophysical observations.
\end{abstract}

\keywords{\href{https://vocabs.ardc.edu.au/repository/api/lda/aas/the-unified-astronomy-thesaurus/current/resource.html?uri=http://astrothesaurus.org/uat/304}{Core-collapse supernovae (304)}; \href{https://vocabs.ardc.edu.au/repository/api/lda/aas/the-unified-astronomy-thesaurus/current/resource.html?uri=http://astrothesaurus.org/uat/1108}{Neutron stars (1108)}}

\section{Introduction}
\label{sec:1}

The equation of state (EOS) of hot and dense matter is an essential ingredient
for understanding astrophysical phenomena, such as core-collapse supernovae,
proto-neutron star cooling, and binary neutron star mergers~\citep{oert17}.
The EOS should cover a wide range of temperatures $T$, proton fractions $Y_p$,
and baryon mass densities $\rho_B$, exhibiting a complex phase diagram.
At low temperatures and subsaturation densities, the matter is nonuniform,
with heavy nuclei forming to lower the free energy of the system.
When the density reaches about half of the nuclear saturation density,
heavy nuclei tend to dissolve into a homogeneous nuclear liquid.
At the density higher than two to three times the nuclear saturation density,
non-nucleonic degrees of freedom, such as hyperons and quarks, may occur
and soften the EOS of dense matter~\citep{maru07,yasu14,webe19,huang22a,huang22b,sumi23}.
On the other hand, with increasing temperature $T$, the density range
of nonuniform matter shrinks and, finally, heavy nuclei cannot be formed
above a critical value of $\sim 14$ MeV~\citep{shen11}.

Owing to the complexity of the phase behavior of stellar matter,
constructing a full EOS for general usage in astrophysical applications
is a daunting task. A set of the available EOSs has been summarized
in the review by~\citet{oert17}, which can also be obtained
from the public database CompOSE~\citep{type22}.
Over the past decades, the two most commonly used EOSs in astrophysical
simulations have been the EOS of~\citet{latt91} and that of~\citet{shen98b}.
The Lattimer-Swesty EOS was based on the compressible liquid drop (CLD) model
with a nonrelativistic Skyrme force. In contrast, the Shen EOS employed the
relativistic mean-field (RMF) model and the Thomas-Fermi approximation
with a parameterized nucleon distribution for the description of nonuniform matter.
In these EOSs, the single-nucleus approximation (SNA) was adopted,
which meant that only a single representative nucleus was considered instead of
an ensemble of nuclei. Recently, EOS tables have been developed beyond the SNA,
by including multiple nuclei in nuclear statistical equilibrium (NSE)~\citep{hemp10,shenG11a,stei13,furu17a,schn17,radu19}.
It has been shown that considering the nuclear distributions may play an important
role in the neutrino-matter interactions~\citep{naga19}, but have less influence
on the thermodynamic quantities of dense matter~\citep{hemp10}.
Furthermore, microscopic approaches based on realistic nuclear forces have also been
used to construct the EOS table for astrophysical simulations~\citep{toga17,furu17b,furu20}.

The first version of the Shen EOS (EOS1) was published in~\citet{shen98b}, in which we
first provided the EOS in three-dimensional ($T$, $Y_p$, $\rho_B$) tabular form,
including the thermodynamic and compositional quantities needed in the applications.
This EOS design is convenient for performing supernova simulations and has been
commonly used in building EOS tables in subsequent years~\citep{type15}.
In~\citet{shen11}, we recalculated the EOS table with improved designs of ranges
and grids, according to the requirements of the EOS users. The improved EOS tables
were referred to as: (a) EOS2, which includes only the nucleonic degree of freedom;
and (b) EOS3, which incorporates additional $\Lambda$ hyperons.
Both EOS1 and EOS2 were based on the RMF approach, using the TM1 parameterization for
nuclear interactions. The nonuniform matter, consisting of a lattice of
heavy nuclei, was described within the Thomas-Fermi approximation, in combination
with assumed nucleon distribution functions and a free-energy minimization procedure.
The TM1 model provides a satisfactory description for finite nuclei and predicts
a maximum neutron star mass of $2.18\ M_\odot$ with nucleonic degrees of freedom
only. However, the resulting neutron star radii seem to be excessively  large~\citep{shen98a,shen20}.
Remarkable progress in astrophysical observations has been achieved over the last
decade, providing crucial constraints on the EOS of dense matter.
A stringent constraint comes from the precise mass measurements of massive pulsars\textemdash
PSR J1614-2230 ($1.908\,\pm\, 0.016 \ M_\odot$;~\citet{arzo18}),
PSR J0348+0432 ($2.01\, \pm 0.04 \, \ M_\odot$;~\citet{anto13}), and
PSR J0740+6620 ($2.08 \, \pm 0.07 \, \ M_\odot$;~\citet{fons21})\textemdash which require the predicted maximum neutron star mass to be larger than
$\sim 2\ M_\odot$.
The first detection of gravitational waves from a binary neutron star merger,
GW170817~\citep{abbo17,abbo18} provided an upper limit on the tidal deformability and constrained the radius of neutron stars~\citep{fatt18,most18}.
Furthermore, the recent observations by the Neutron Star Interior Composition Explorer (NICER)
for PSR J0030+0451~\citep{mill19,rile19} and PSR J0740+6620~\citep{mill21,rile21}
provided simultaneous measurements of the mass and radius of neutron stars,
which offer strong constraints on the EOS of dense matter.
Considering the progress in astrophysical observations, we constructed a revised
version of the Shen EOS (EOS4), based on an extended TM1 model, referred to as the
TM1e model, in~\citet{shen20}.
It is noteworthy that the TM1e and original TM1 models have identical properties
for symmetric nuclear matter but exhibit different behaviors of the symmetry energy.
The TM1e model has a symmetry energy slope of $L=40$ MeV, significantly
smaller than the value of $L=111$ MeV in the original TM1 model.
Consequently, it predicts smaller neutron star radii, which are supported by
astrophysical observations~\citep{shen20}.
By comparing the results from astrophysical simulations using EOS4 and EOS2,
one can estimate the effects of the symmetry energy and its density dependence~\citep{sumi19}.

Recently, it was reported in~\citet{schn19} and \citet{yasi20} that the effective nucleon mass
has a decisive effect on supernova explosions through pressure difference,
proto-neutron star contraction, and neutrino emission (see also~\citet{ande21}
for gravitational waves).
A larger effective mass leads to smaller thermal contributions to the pressure,
which results in more rapid contraction and aids the shock evolution in faster explosions. However, in their calculations, the influence of the effective mass was
investigated using a set of Skyrme-type EOSs, by varying the effective nucleon
mass at saturation density, which was treated as a model parameter within a
nonrelativistic framework.
In~\citet{naka19}, the influence of the effective nucleon mass on the cooling process
of a proto-neutron star was investigated using a series of phenomenological EOSs,
in which the effective nucleon mass, as a model parameter, was set to be constant.
It is widely accepted that the effective mass should be density-dependent and can be incorporated more consistently in a relativistic framework than in nonrelativistic approaches.

In the present work, we will adjust the RMF parameters based on the TM1e
parameterization, so that the saturation properties obtained by the new
parameterization remain the same as those of TM1e, but with different effective masses.
The TM1e model predicts an effective mass ratio $M^{\ast}/M \sim 0.63$ at the
saturation density $n_0$, while the new parameterization, referred to hereafter
as the TM1m model, sets this ratio to be $M^{\ast}/M \sim 0.8$.
Note that the effective masses in the TM1e and TM1 models are identical, 
due to having the same isoscalar properties.

Currently, there is no direct constraint on the effective nucleon
mass from experimental observations. In nonrelativistic Skyrme models,
the effective mass at saturation density is often treated as a model parameter,
which can be tuned to different values within the range of $M^{\ast}/M \sim 0.5-1.0$,
while keeping other saturation properties unchanged~\citep{naka19,schn19}.
In RMF models, the effective mass, defined by $M^{\ast}=M+g_{\sigma}\sigma$,
is reduced from the free nucleon mass by a large negative scalar
potential, $U_S=g_{\sigma}\sigma$.
The saturation mechanism of symmetric nuclear matter
in the RMF models can be achieved by a delicate
balance between large negative scalar and positive vector potentials~\citep{sero86}.
As a consequence, the effective mass at saturation density obtained in
well-known RMF models~\citep{sero86,lala97,stei13,chen14,dutr14,miya24} generally
lies within a narrow range, $M^{\ast}/M \sim 0.54-0.8$.
In~\citet{choi21}, the authors investigated the impacts of the effective nucleon mass
on the nuclear and neutron star properties within the range
of $0.5\leq M^{\ast}/M \leq 0.8$.
It is almost impossible to enforce a very large effective mass in the RMF model,
because small scalar and vector potentials, which are related to large effective masses,
will not be able to reproduce the empirical saturation point of symmetric nuclear matter.
Therefore, we prefer to set $M^{\ast}/M \sim 0.8$ in the TM1m model
and compare to the TM1e model with $M^{\ast}/M \sim 0.63$, while
keeping other saturation properties unchanged.

In order to investigate the impact of the effective nucleon mass on astrophysical
simulations, we construct a new EOS table (referred to as EOS5) using the TM1m model.
We prepare all quantities in the EOS table for uniform matter at densities
higher than $\sim 10^{14}\,\mathrm{g/cm^{3}}$ by using the TM1m model,
which can then be combined with the nonuniform part of the Shen EOS4 at low densities
to generate the final EOS table, as is often done in the literature~\citet{ishi08,sumi19}.
By switching only the uniform matter in EOS5 with the TM1m model, we clarify
the influence of effective mass at high densities.
The application of EOS5 with $M^{\ast}/M \sim 0.8$, compared to EOS4
with $M^{\ast}/M \sim 0.63$, can be used to clarify the impact of effective mass
on astrophysical simulations, such as core-collapse supernovae, proto-neutron
star cooling, and binary neutron star mergers.

This paper is arranged as follows. In Section~\ref{sec:2}, we briefly
describe the RMF approach used for constructing the EOS table
and compare different parameter sets.
In Section~\ref{sec:3}, we present numerical results using the new
parameterization TM1m and compare them to those obtained from other models,
including both relativistic and nonrelativistic approaches.
Finally, a summary and conclusions are presented in Section~\ref{sec:4}.

\begin{table*}[tbp]
	\caption{Coupling constants of the TM1m, TM1e, and original TM1 models.}
	\begin{center}
		\begin{tabular}{p{1.6cm}<{\raggedright}p{2cm}<{\centering}p{2cm}<{\centering}p{2cm}<{\centering}p{2cm}<{\centering}p{1.8cm}<{\centering}p{1.8cm}<{\centering}p{1.8cm}<{\centering}}
			\hline\hline
			Model & $g_\sigma$  & $g_\omega$ & $g_\rho$ & $g_{2}$ [fm$^{-1}$] & $g_{3}$ & $c_{3}$ & $\Lambda_{\textrm{v}}$ \\
			\hline
			TM1m  & 7.93528     & 8.63169    & 11.51296  & $-$11.51628     &54.88715 & 0.00025 &
			0.09326 \\
			TM1e  & 10.0289     & 12.6139    & 13.9714  & $-$7.2325        &0.6183   & 71.3075 & 0.0429  \\
			TM1   & 10.0289     & 12.6139    &  9.2644  & $-$7.2325        &0.6183   & 71.3075 & 0.0000  \\
			\hline\hline
		\end{tabular}
		\label{tab:1}
	\end{center}
\end{table*}

\begin{table}[tbp]
	\begin{center}
		\caption{Nuclear matter properties obtained in the TM1m, TM1e, and original TM1 models.}
		\begin{footnotesize}
			\begin{tabular}{p{1.75cm}<{\raggedright}p{0.7cm}<{\centering}p{0.7cm}<{\centering}p{0.7cm}<{\centering}p{0.7cm}<{\centering}p{0.7cm}<{\centering}p{0.75cm}<{\centering}}
			\hline\hline
			Model& $n_0$   & $\text{E/A}$   & $K$     &$E_{\mathrm{sym}}$& $L$     &$M^{\ast}/M$ \\
			&(fm$^{-3})$ & (MeV) & (MeV) & (MeV) & (MeV)   &
			\\  \hline
			TM1m         & 0.145   & -16.3   & 281     & 31.4       & 40      & 0.793  \\
			TM1e         & 0.145   & -16.3   & 281     & 31.4       & 40      & 0.634  \\
			TM1          & 0.145   & -16.3   & 281     & 36.9       & 111     & 0.634  \\
			SFHo         & 0.158   & -16.2   & 245     & 31.6       & 47      & 0.761  \\
			Skyrme-SLy4   & 0.160   & -16.0   & 230     & 32.0       & 46      & 0.694  \\
			\hline\hline
		\end{tabular}	
		\end{footnotesize}
		\label{tab:2}
	\end{center}
\textbf{Note.} The saturation density and the energy per particle are denoted by $n_0$ and $\text{E/A}$, the incompressibility by $K$, the symmetry energy and its slope by $E_{\mathrm{sym}}$ and $L$, and the effective mass ratio by $M^{\ast}/M$. The results obtained from the SFHo model~\citep{stei13} and the	nonrelativistic Skyrme-SLy4 model~\citep{chab98} are also listed for comparison.
\end{table}

\section{Model and parameters}
\label{sec:2}

In order to make the article self-contained, we give a brief description of
the RMF model used for constructing the EOS table.
We employ the RMF model, including nonlinear terms for the $\sigma$ and $\omega$ mesons
and an additional $\omega$-$\rho$ coupling term. Nucleons interact through
the exchange of mesons\textemdash specifically, scalar $\sigma$, vector $\omega$,
and isovector $\rho$ mesons~\citep{bao14b,shen20}.
The nucleonic Lagrangian density can be expressed as
\begin{eqnarray}
\label{eq:LRMF}
\mathcal{L}_{\mathrm{RMF}} & = & \sum_{i=p,n}\bar{\psi}_i
\left[ i\gamma_{\mu}\partial^{\mu}-\left(M+g_{\sigma}\sigma\right) \right. \nonumber \\
&& \left. -\gamma_{\mu} \left(g_{\omega}\omega^{\mu} +\frac{g_{\rho}}{2}
\tau_a\rho^{a\mu}\right)\right]\psi_i  \nonumber \\
&& +\frac{1}{2}\partial_{\mu}\sigma\partial^{\mu}\sigma
-\frac{1}{2}m^2_{\sigma}\sigma^2-\frac{1}{3}g_{2}\sigma^{3} -\frac{1}{4}g_{3}\sigma^{4}
\nonumber \\
&& -\frac{1}{4}W_{\mu\nu}W^{\mu\nu} +\frac{1}{2}m^2_{\omega}\omega_{\mu}\omega^{\mu}
+\frac{1}{4}c_{3}\left(\omega_{\mu}\omega^{\mu}\right)^2  \nonumber
\\
&& -\frac{1}{4}R^a_{\mu\nu}R^{a\mu\nu} +\frac{1}{2}m^2_{\rho}\rho^a_{\mu}\rho^{a\mu} \nonumber \\
&& +\Lambda_{\mathrm{v}} \left(g_{\omega}^2 \omega_{\mu}\omega^{\mu}\right)
\left(g_{\rho}^2\rho^a_{\mu}\rho^{a\mu}\right),
\end{eqnarray}
where $W^{\mu\nu}$ and $R^{a\mu\nu}$ represent the antisymmetric field
tensors corresponding to $\omega^{\mu}$ and $\rho^{a\mu}$, respectively.
Within the mean-field approximation, the meson field operators are replaced
by their expectation values.
In a static uniform system, the nonzero components are given by
$\sigma =\left\langle \sigma \right\rangle$, $\omega =\left\langle
\omega^{0}\right\rangle$, and $\rho =\left\langle \rho^{30} \right\rangle$.
The equations of motion for nucleons and mesons are derived from the Lagrangian
density, and these coupled equations can be solved self-consistently within the RMF framework.

In order to explore the influence of the effective nucleon mass on astrophysical
simulations, we refit the RMF parameters based on the TM1e model used in
constructing the Shen EOS4~\citep{shen20}.
A new parameter set, referred to as TM1m, is introduced with a larger
effective mass, while all other saturation properties remain identical
to those of TM1e.
In Table~\ref{tab:1}, we present the coupling constants of the TM1m, TM1e,
and original TM1 models, while the corresponding saturation properties are listed
in Table~\ref{tab:2}.
It is shown that the main difference between TM1m and TM1e
is the effective mass, i.e., $M^{\ast}/M \sim 0.8$ in the TM1m model and
$M^{\ast}/M \sim 0.63$ in the TM1e model. On the other hand, the difference
between TM1e and TM1 lies in the density dependence of the symmetry energy, with
the symmetry energy slope being $L=40$ MeV in the TM1e model and $L=111$ MeV
in the original TM1 model, respectively.

We prefer to set $M^{\ast}/M \sim 0.8$ at saturation density
in the TM1m model, which is considered to be a relatively high and achievable
value within the RMF approach~\citep{dutr14}.
It is noteworthy that the allowed region for the ratio $M^{\ast}/M$ in the RMF model
is significantly restricted compared to nonrelativistic Skyrme models.
This is because the saturation mechanism of symmetric nuclear matter
in the RMF approach is due to
a delicate balance between large negative scalar and positive vector potentials.
An excessively large effective mass $M^{\ast}$ correlates with scalar and vector potentials that are too weak to accurately reproduce the empirical saturation point.
In fact, the effective mass at saturation density obtained in several well-known
RMF models lies within a narrow range.
For example, the linear Walecka model~\citep{sero86} has $M^{\ast}/M = 0.54$,
while this ratio is around 0.6 in the nonlinear RMF models
with NL3~\citep{lala97}, IUFSU~\citep{fatt10}, FSUGold2~\citep{chen14}, and FSUGarnet~\citep{chen15} parameterizations.
For comparison, we include in Table~\ref{tab:2} the results from
an RMF model with the SFHo parameterization~\citep{stei13}
and a nonrelativistic Skyrme model with the SLy4 parameterization~\citep{chab98}.
The SFHo model has been widely used in astrophysical simulations~\citep{oert17},
in which $M^{\ast}/M = 0.76$ is obtained and several additional couplings
are introduced~\citep{stei13}.
As a typical Skyrme force, the SLy4 model is frequently utilized for
both nuclear structure calculations and astrophysical simulations~\citep{oert17,schn17},
employing a zero-range effective interaction within the nonrelativistic framework.
We intend to compare the results obtained from the RMF models with those from
the Skyrme-SLy4 model to explore possible relativistic effects,
particularly at high densities.

\begin{figure}[htbp]
	\centering
	\includegraphics[width=8.5 cm]{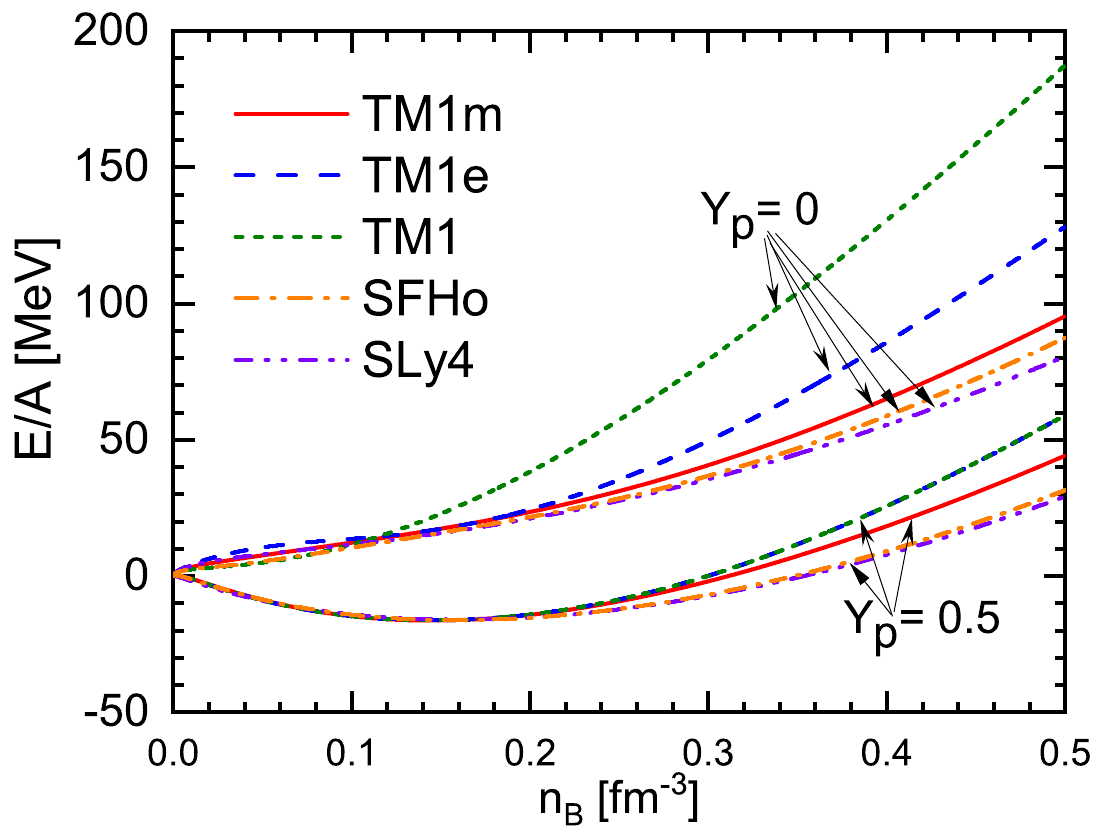}
	\caption{Energy per baryon $\text{E/A}$ of symmetric nuclear matter and neutron matter
		as a function of the baryon number density $n_B$.}
	\label{fig:21EA}
\end{figure}

In Figure~\ref{fig:21EA}, we plot the energy per baryon $\text{E/A}$ of symmetric nuclear
matter and neutron matter as a function of the baryon number density $n_B$.
It is shown that the behaviors of symmetric nuclear matter are exactly the same
between the TM1e and TM1 models, whereas those of the TM1m model become slightly
lower as the density increases. This is because the larger effective mass in the
TM1m model leads to a smaller kinetic energy.
Moreover, the TM1m model exhibits less attraction by $\sigma$ mesons and weaker repulsion
by $\omega$ mesons, compared to the TM1e and TM1 models.
Relativistic effects are well known to increase with density, which can yield
relatively smaller kinetic energy for particles with larger mass, especially
at higher densities.

The curve of the SFHo model for $Y_p=0.5$ is lower than
that of TM1m. This is mainly because the incompressibility in the SFHo
model ($K=245$ MeV) is significantly smaller than that in the TM1m
model ($K=281$ MeV), despite its effective mass being even smaller.
It is well known that a small value of $K$ results in a slowly rising $\text{E/A}$
curve for symmetric nuclear matter as the density increases.
On the other hand, the results of the SLy4 model are very close to that of SFHo.
Comparing these two models, the nonrelativistic Skyrme-SLy4 model has
$M^{\ast}/M = 0.69$, which is smaller than the $M^{\ast}/M = 0.76$ of the SFHo model
and is helpful for raising the $\text{E/A}$ curve. However, a slightly larger $K$ and
relativistic effects in the SFHo model almost balance out the discrepancy
induced by $M^{\ast}/M$. Ultimately, both the SFHo and SLy4 models yield
remarkably similar $\text{E/A}$ curves for symmetric nuclear matter.
In the case of neutron matter ($Y_p=0$), significant differences are observed
between these models, which are caused by the differences in the symmetry energy
slope $L$ and relativistic effects. For example, the original TM1 model
has a rather large value of $L=111$ MeV, which leads to a rapidly rising $\text{E/A}$.
In contrast, all other models considered here have similar small values of $L$,
so the differences between them are likely attributed to different $M^{\ast}/M$
and relativistic effects.

\begin{figure}[htbp]
	\centering
	\includegraphics[width=8.5 cm]{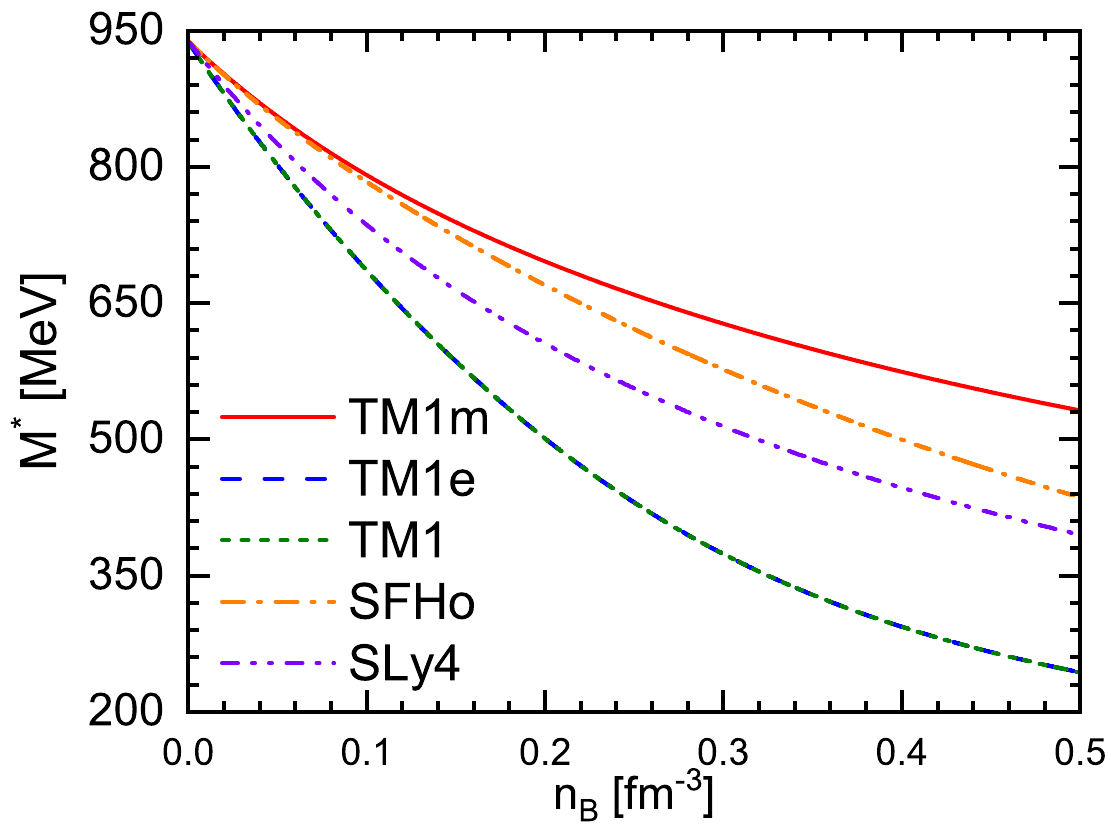}
	\caption{Effective nucleon mass $M^{\ast}$ as a function of the baryon number
		density $n_{B}$.}
	\label{fig:22Mn}
\end{figure}

In Figure~\ref{fig:22Mn}, we show the effective nucleon mass $M^{\ast}$ as a function
of the baryon number density $n_{B}$. It is seen that the TM1e model maintains the
exact same behavior for $M^{\ast}$ as the original TM1 model,
whereas $M^{\ast}$ in the TM1m model is notably larger than that in TM1.

The effective masses obtained from the SFHo and SLy4 models
are also depicted for comparison, falling between those of the TM1m and TM1e models.
The effective mass is known to play a key role
in neutrino emission processes~\citep{yasi20}, motivating us to explore the effects
caused by the effective mass on the realistic EOS table.
On the other hand, the effective mass is closely related to the spin-orbit splittings
in finite nuclei\textemdash namely, a larger Dirac mass corresponds to a
smaller spin-orbit splitting.
The behavior of spin-orbit splittings can be improved
by introducing an additional tensor interaction, which does not contribute to
uniform matter at high densities~\citep{type20}.

\begin{figure}[htbp]
	\centering
	\includegraphics[width=8.5 cm]{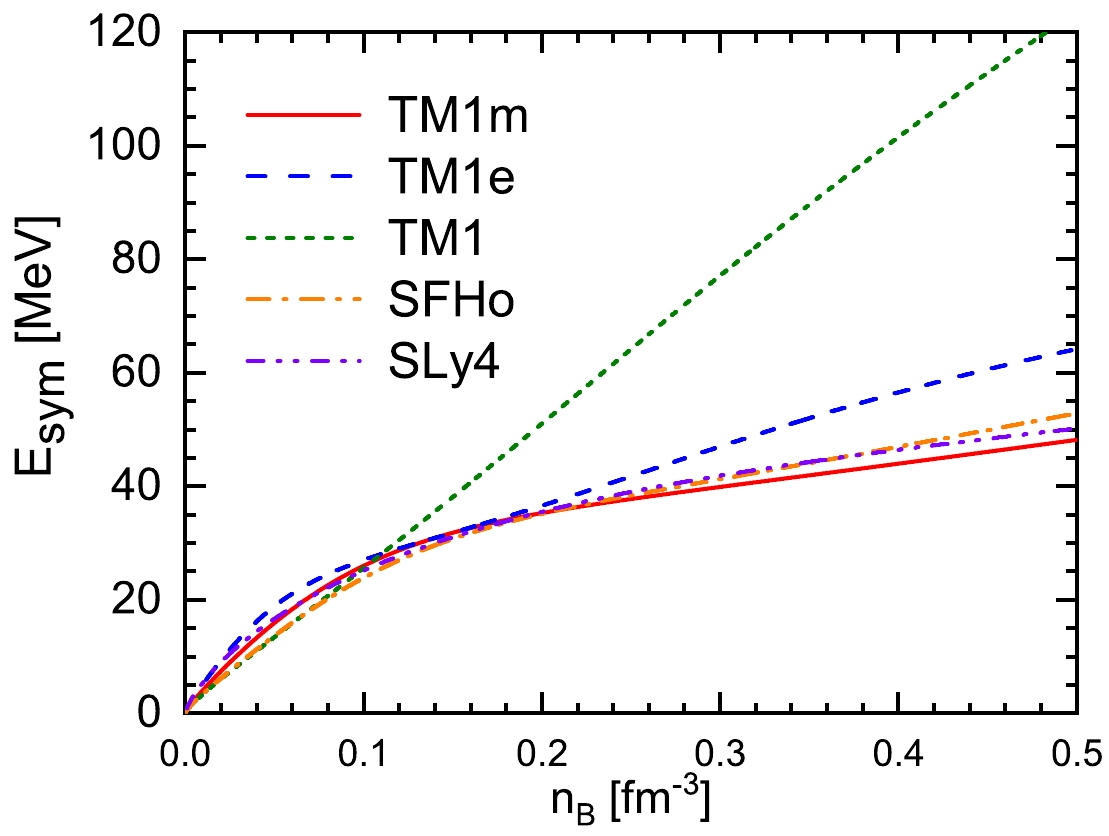}
	\caption{Symmetry energy $E_{\mathrm{sym}}$ as a function of the baryon number
		density $n_{B}$.}
	\label{fig:23Esym}
\end{figure}

In Figure~\ref{fig:23Esym}, we display the symmetry energy $E_{\mathrm{sym}}$ as a function of the baryon number density $n_{B}$. At higher densities, $E_{\mathrm{sym}}$
in the TM1e model is lower than that in the original TM1 model.
This is because the TM1e model has a relatively smaller slope parameter ($L=40$ MeV)
than the TM1 model ($L=111$ MeV).
On the other hand, $E_{\mathrm{sym}}$ in the TM1m model is slightly lower than that
in the TM1e model, which is more evident with increasing density.
Although the TM1m model has the same slope parameter ($L=40$ MeV) as the TM1e model,
its larger effective mass, together with the smaller coupling constants $g_\sigma$ and $g_\omega$, can reduce the relativistic effects and lead to smaller
$E_{\mathrm{sym}}$ at higher densities. It is noteworthy that these three models
have the same values of $E_{\mathrm{sym}}$ at a density of $0.11\, \mathrm{fm}^{-3}$.
This is because the model parameters are chosen to keep $E_{\mathrm{sym}}$ fixed
at the density of $0.11\, \mathrm{fm}^{-3}$. This choice is due to the fact
that the binding energies of finite nuclei are essentially determined by the
symmetry energy at a density of $0.11\, \mathrm{fm}^{-3}$, not by the symmetry
energy at saturation density~\citep{zhang13,bao14b}.

It is seen that the values of $E_{\mathrm{sym}}$ in the SFHo
and SLy4 models are very close to those in the TM1m model.
This is because these models have similar slope parameters $L$
and effective masses $M^{\ast}$, as shown in Table~\ref{tab:2}.

To construct a realistic EOS table for general usage in astrophysical simulations,
we should perform calculations covering a wide range of temperatures $T$,
proton fractions $Y_p$, and baryon mass densities $\rho_B$.
For uniform nuclear matter at densities above $\sim 10^{14}\,\mathrm{g/cm^{3}}$,
all required quantities can be derived within the RMF framework.
To ensure completeness, we present the key thermodynamic quantities of uniform
nuclear matter: the energy density $\epsilon$, entropy density $s$, and pressure $p$.
The energy density in the TM1m model is given by
\begin{eqnarray}
\label{eq:ERMF}
\epsilon &=& \displaystyle{\sum_{i=p,n} \frac{1}{\pi^2}
  \int_0^{\infty} dk\,k^2\,
  \sqrt{k^2+{M^{\ast}}^2}\left( f_{i+}^{k}+f_{i-}^{k}\right) } \nonumber\\
 & &
  +\frac{1}{2}m_{\sigma}^2\sigma^2+\frac{1}{3}g_{2}\sigma^{3}
  +\frac{1}{4}g_{3}\sigma^{4}
  +\frac{1}{2}m_{\omega}^2\omega^2 \nonumber\\
 & &
  +\frac{3}{4}c_{3}\omega^{4}+\frac{1}{2}m_{\rho}^2\rho^2
  +3 \Lambda_{\mathrm{v}}\left(g^2_{\omega}\omega^2\right)
     \left(g^2_{\rho}\rho^2\right),
\end{eqnarray}
the entropy density is written as
\begin{eqnarray}
	\label{eq:SRMF}
	s &=& -\displaystyle{\sum_{i=p,n}\frac{1}{\pi^{2}}
		\int_{0}^{\infty}dk\,k^{2}}
	\left[ f_{i+}^{k}\ln f_{i+}^{k} \right. \nonumber \\
	& &  +\left( 1-f_{i+}^{k}\right)
	\ln \left(1-f_{i+}^{k}\right)   \nonumber \\
	& & \left. +f_{i-}^{k}\ln f_{i-}^{k}
	+\left( 1-f_{i-}^{k}\right) \ln \left( 1-f_{i-}^{k}\right) \right],
\end{eqnarray}
and the pressure is given by
\begin{eqnarray}
\label{eq:PRMF}
 p &=& \displaystyle{\sum_{i=p,n} \frac{1}{3\pi^2}
   \int_0^{\infty} dk\,k^2\,
   \frac{k^2}{\sqrt{k^2+{M^{\ast}}^2}}
   \left( f_{i+}^{k}+f_{i-}^{k}\right) } \nonumber\\
 & &
  -\frac{1}{2}m_{\sigma}^2\sigma^2-\frac{1}{3}g_{2}\sigma^{3}
  -\frac{1}{4}g_{3}\sigma^{4}+\frac{1}{2}m_{\omega}^2\omega^2 \nonumber\\
 & &
  +\frac{1}{4}c_{3}\omega^{4}
  +\frac{1}{2}m_{\rho}^2\rho^2
  +\Lambda_{\rm{v}}\left(g^2_{\omega}\omega^2\right)
   \left(g^2_{\rho}\rho^2\right).
\end{eqnarray}
Here, $M^{\ast}=M+g_{\sigma}\sigma$ denotes the effective nucleon mass.
$f_{i+}^{k}$ and $f_{i-}^{k}$ ($i=p,n$) are the occupation
probabilities of nucleons and antinucleons at momentum $k$, respectively, 
which are given by the Fermi-Dirac distribution:
\begin{eqnarray}
\label{eq:firmf}
f_{i\pm}^{k}=\left\{1+\exp \left[ \left( \sqrt{k^{2}+{M^{\ast}}^2}
  \mp \nu_{i}\right)/T\right]
 \right\}^{-1}.
\end{eqnarray}
The kinetic part of the chemical potential $\nu_i$ is related to
the chemical potential $\mu_i$ as
\begin{eqnarray}
\mu_{i} = \nu_{i} + g_{\omega}\omega + \frac{g_{\rho}}{2}\tau_{3i}\rho.
\end{eqnarray}
The number density of protons ($i=p$) or neutrons ($i=n$) can be obtained by
\begin{equation}
\label{eq:nirmf}
 n_{i}=\frac{1}{\pi^2}
       \int_0^{\infty} dk\,k^2\,\left(f_{i+}^{k}-f_{i-}^{k}\right).
\end{equation}

The free-energy density is given by $f=\epsilon - Ts$.
Within the RMF model for uniform matter, these thermodynamic quantities satisfy
the self-consistent relation
\begin{equation}
\label{eq:thermo}
f = \sum_{i=p,n} \mu_i n_i - p.
\end{equation}

\section{Results and discussion}
\label{sec:3}

In this section, we present numerical results using the TM1m model with a larger
effective mass, while also conducting an in-depth comparison with the TM1e and
original TM1 models. First, we discuss the properties of neutron stars at zero
temperature and check the compatibility with current observations.
Second, we show the results of the new EOS table (EOS5), especially at densities
higher than $\sim 10^{14}\,\mathrm{g/cm^{3}}$, in order to explore the impact of the 
effective nucleon mass on the EOS table.

\subsection{Neutron stars at zero temperature}
\label{sec:3.1}

\begin{figure}[htbp]
 \begin{center}
  \includegraphics[clip,width=8.5 cm]{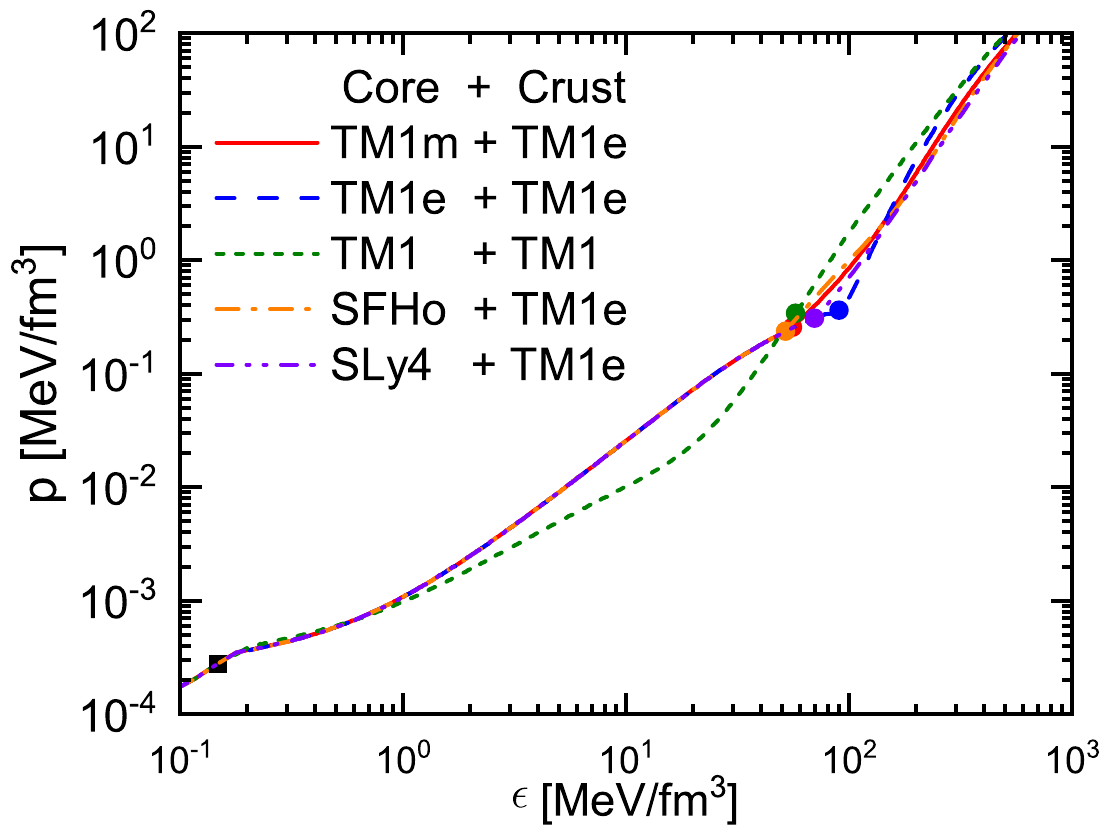}
  \caption{Pressure $p$ as a function of the energy density $\epsilon$.
  The core EOSs are obtained within the models considered in this work.
  The inner crust is described in the Thomas-Fermi approximation using
  the TM1e and TM1 models. The crust-core transition is indicated by
  the filled circles. The BPS EOS is adopted for the outer crust and
  the matching point is marked by the filled square. }
  \label{fig:31PE0}
 \end{center}
\end{figure}
\begin{figure}[htbp]
 \begin{center}
  \includegraphics[clip,width=8.5 cm]{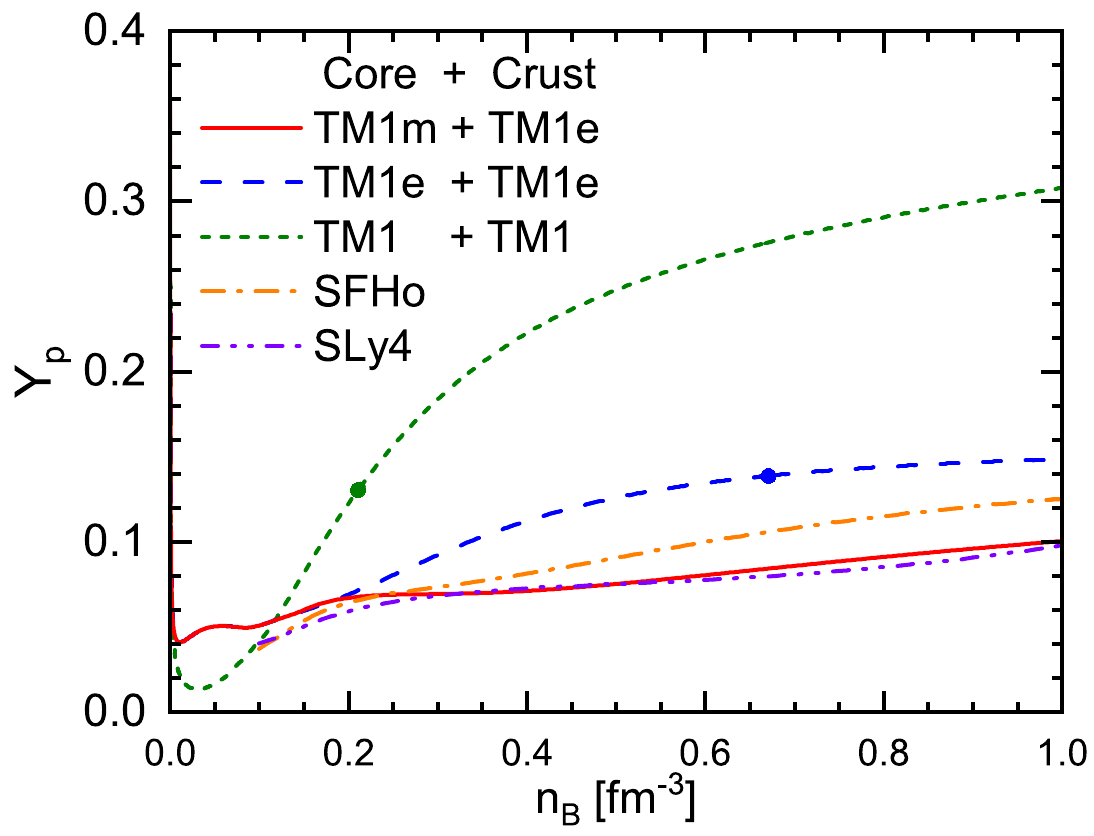}
  \caption{Proton fraction $Y_p$ as a function of the baryon
  density $n_B$. The filled circles indicate the threshold for the DUrca process.}
  \label{fig:32YP0}
 \end{center}
\end{figure}
\begin{figure}[htbp]
 \begin{center}
  \includegraphics[clip,width=8.5 cm]{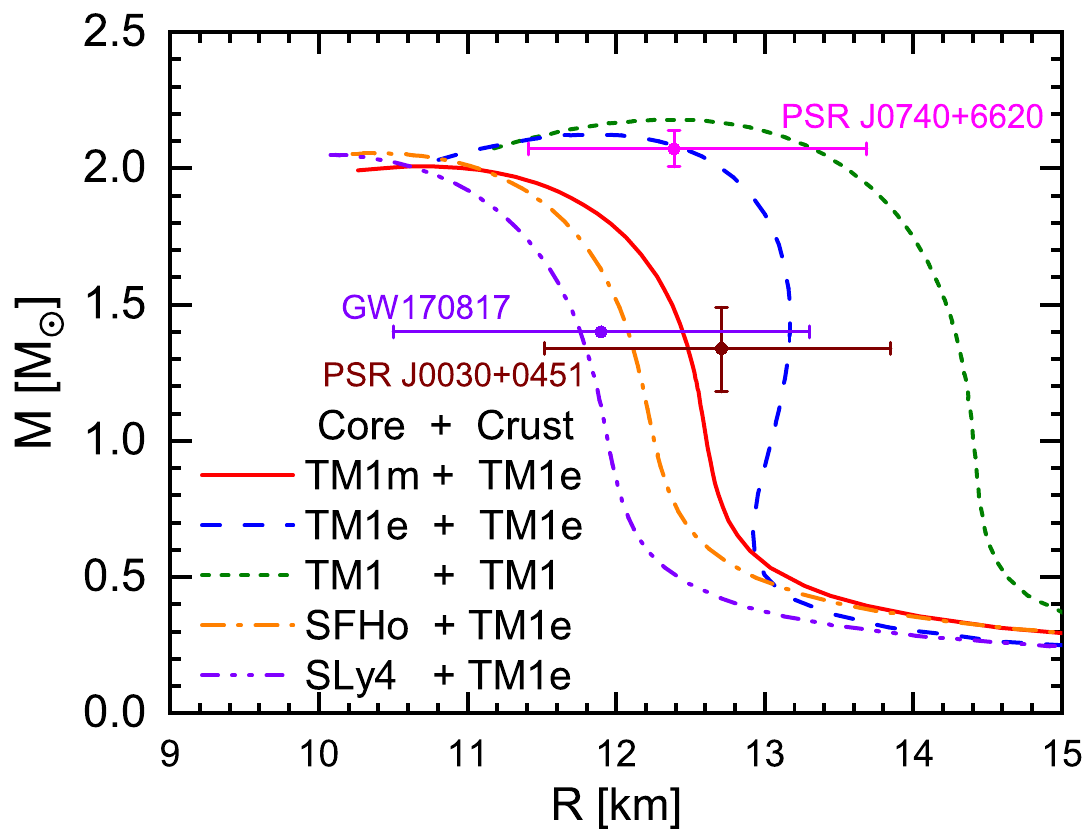}
  \caption{Mass-radius relations of neutron stars obtained using the
  EOSs shown in Figure~\ref{fig:31PE0}.
  The horizontal violet line represents the constraint on $R_{1.4}$
  inferred from GW170817~\citep{abbo18}.
  The horizontal brown and magenta lines correspond to simultaneous measurements
  of the mass and radius from NICER for PSR J0030+0451~\citep{rile19} and
  PSR J0740+6620~\citep{rile21}, respectively. }
  \label{fig:33MR0}
 \end{center}
\end{figure}
\begin{figure*}[htbp]
  \includegraphics[clip,width=8.5 cm]{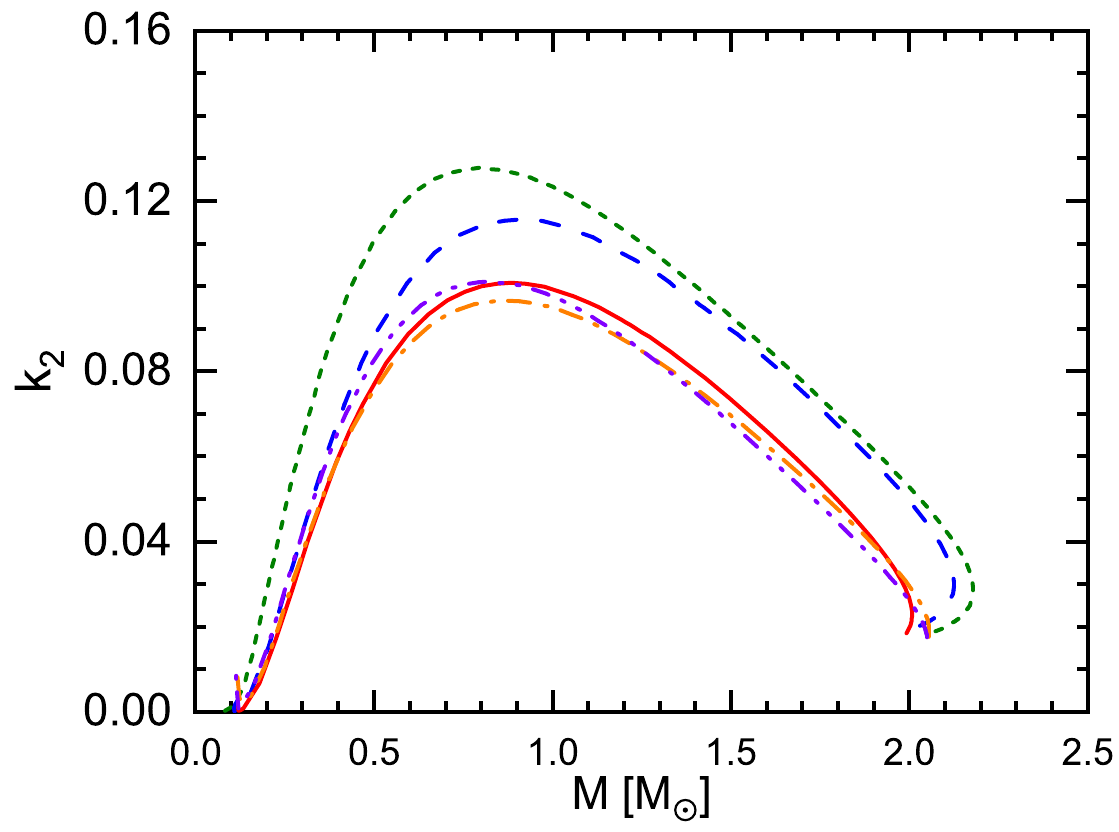}
  \includegraphics[clip,width=8.5 cm]{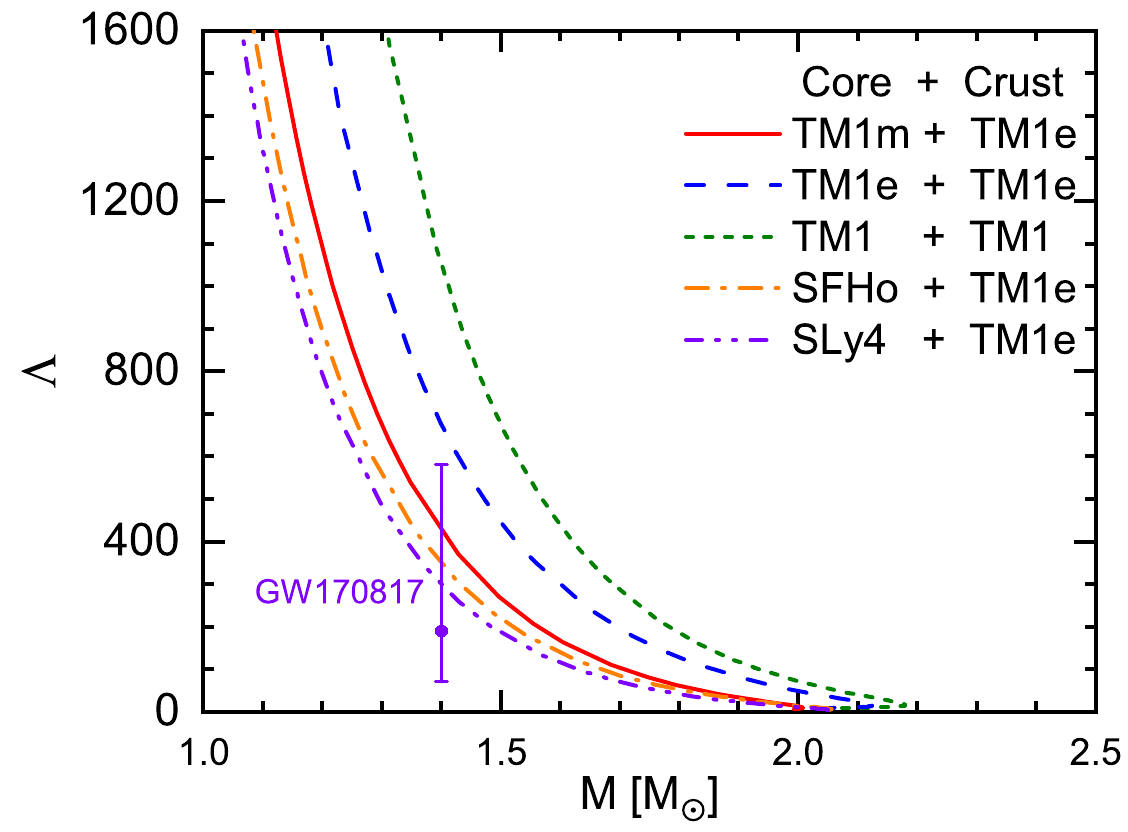}
  \caption{Love number $k_{2}$ and tidal deformability $\Lambda$
  as a function of the neutron star mass $M$.
  The vertical violet line represents the constraint on $\Lambda_{1.4}$
  inferred from GW170817~\citep{abbo18}.}
\label{fig:34KL0}
\end{figure*}

The properties of static neutron stars can be obtained by solving the well-known
Tolman-Oppenheimer-Volkoff equation with the EOS over a wide range of densities.
The neutron star matter is assumed to be in beta equilibrium with the charge neutrality
at zero temperature. Generally, the EOS used for the calculations of the neutron star
structure can be divided into at least three segments:
(a) the EOS of the outer crust below the neutron drip density;
(b) the EOS of the inner crust from the neutron drip to crust-core transition; and
(c) the EOS of the liquid core above the crust-core transition.
In the present work, we use the Baym-Pethick-Sutherland (BPS) EOS~\citep{baym71}
for the outer crust, while the inner crust EOS is based on the self-consistent
Thomas-Fermi approximation using both TM1e ($L=40$ MeV) and TM1 ($L=111$ MeV) parameterizations~\citep{ji19}.

The EOS of the liquid core above the crust-core transition is calculated in
the RMF approach using the TM1m, TM1e, and TM1 parameterizations, while the results
of the SFHo and SLy4 models are also depicted for comparison.
The core EOS is connected to the inner-crust EOS at their crossing point.
In Figure~\ref{fig:31PE0}, we show the pressure $p$ as a function of the energy
density $\epsilon$. It is shown that the TM1m model predicts relatively small
pressures at high densities, while the TM1 EOS is stiffer than the other cases.
We note that the EOSs with TM1e (blue dashed line) and TM1 (green dotted line)
are unified EOSs, because their inner-crust and core segments are obtained
within the same nuclear model.
It is seen that the core EOSs obtained from the SFHo and SLy4 models are
very close to the TM1m model.

In Figure~\ref{fig:32YP0}, the proton fraction $Y_p$ is plotted as a function of
the baryon density $n_B$, and the corresponding threshold density for the DUrca process
is indicated by the filled circles. It is well known that the DUrca process can occur
for $Y_p \geq 1/9$ in a simple $npe$ matter, while the critical $Y_p$ for the DUrca process is in the range of $(11.1-14.8)\%$ when the muons are included under
the equilibrium condition $\mu_e=\mu_\mu$~\citep{latt91b}.
The original TM1 model ($L=111$ MeV) predicts higher $Y_p$ and a smaller
DUrca threshold density ($\sim 0.21\, \rm{fm}^{-3}$).

The results of TM1e are lower than those of TM1,
while its DUrca threshold density is about $0.67\, \rm{fm}^{-3}$.
The curves of TM1m, SFHo, and SLy4 are close to each other and lie lower
than those of TM1 and TM1e. The behaviors of $Y_p$ are mainly determined by the 
symmetry energies, as illustrated in Figure~\ref{fig:23Esym}.
Due to the small symmetry energies in TM1m, SFHo, and SLy4, these models yield
relatively small $Y_p$ in neutron star matter. As a result, the critical $Y_p$
for the DUrca process could not be reached in these models.
It is well known that the DUrca process is the fastest neutrino emission
mechanism ever known. Once it is turned on, the neutron star cooling
is predominantly controlled by the DUrca process, which leads to a rapid
drop in the cooling curve, known as fast cooling.
According to observational data in the temperature-age diagram,
the DUrca process is unlikely to occur in neutron stars with
masses below $1.5\ M_\odot$, since it would lead to an unacceptably
fast cooling, in disagreement with the observations~\citep{fant13}.
On the other hand, for the TM1m, SFHo, and SLy4 models, where the DUrca process
is forbidden, the thermal evolution of neutron stars is regulated by several
slow cooling processes, such as the modified Urca and bremsstrahlung,
which result in higher temperatures than the observations of middle-age
stars~\citep{lim17}. Additional cooling processes, related to the presence of
hyperons and superfluidity, have also been discussed in the literature~\citep{fort21},
but this is beyond the scope of the present paper.

In Figure~\ref{fig:33MR0}, we display the predicted mass-radius relations of
neutron stars using the EOSs shown in Figure~\ref{fig:31PE0}, along with several
constraints from astrophysical observations.
It is found that the maximum masses of neutron stars predicted by the TM1m, TM1e,
and TM1 models are about $2.01$, $2.12$, and $2.18\ M_\odot$, respectively,
which fulfill the observational constraints on the maximum
mass $M_{\mathrm{max}} > 2\ M_\odot$.
The mass-radius relations exhibit significant variations among these models,
which are caused by different behaviors of the symmetry energy and its slope,
as shown in Figure~\ref{fig:23Esym}.
For the radius of a canonical $1.4\ M_\odot$ neutron star, denoted as $R_{1.4}$,
a large value of $\sim 14.2$ km is obtained using the TM1 model, while it
reduces to $13.1$ km in the TM1e model and $12.4$ km in the TM1m model, respectively.

The results obtained from the SFHo and SLy4 models are also
depicted for comparison. The maximum masses of the neutron stars obtained from
the SFHo and SLy4 models are about $2.06\ M_\odot$,
while the values of $R_{1.4}$ are $12.1$ km and $11.7$ km, respectively.
The analysis of GW170817 data provides a constraint on $R_{1.4}$,
with an estimated value of $R_{1.4}=11.9\pm1.4$ km~\citep{abbo18}.
The resulting $R_{1.4}$ within these models, except TM1, can be compatible
with the constraint inferred from GW170817, which is closely related to
the symmetry energy slope $L$.
Generally, a smaller value of $L$ leads to a smaller $R_{1.4}$.

The observation of gravitational waves from GW170817 has provided valuable
insights and constraints on the tidal deformability of neutron stars.
Theoretically, the tidal deformability $\Lambda$ can be calculated using the EOS
through both the tidal Love number $k_2$ and the compactness parameter $C=M/R$,
following the relation $\Lambda=\frac{2}{3}k_2 C^{-5}$~\citep{ji19}.
In Figure~\ref{fig:34KL0}, we show the tidal Love number $k_2$ (left panel)
and the dimensionless tidal deformability $\Lambda$ (right panel) as a
function of the neutron star mass $M$.
It is observed that $k_2$ rises as the neutron star mass increases,
achieving its maximum at a mass of around 0.7 to $0.9\ M_\odot$.
Subsequently, in the more massive range, $k_2$ exhibits a rapid reduction.
It is noticeable that the TM1m model predicts a smaller $k_2$ value
compared to the TM1e and TM1 models, leading to a lower tidal deformability
$\Lambda$ within the TM1m model, as illustrated in the right panel of Figure~\ref{fig:34KL0}.
Generally, the value of $\Lambda$ is very large for a small neutron star mass
due to its small compactness parameter $C$.
As the star mass increases, the tidal deformability $\Lambda$ decreases rapidly.
The analysis of GW170817 has provided a constraint on the tidal deformability of
a $1.4\ M_\odot$ neutron star, i.e., $70 \le \Lambda_{1.4} \le 580$~\citep{abbo18}.
The resulting $\Lambda_{1.4}$ in the TM1m, SFHo, and SLy4 models are
compatible with the constraint inferred from the analysis of GW170817.

\subsection{Supernova matter at finite temperature}
\label{sec:3.2}

To explore the influence of the effective nucleon mass on astrophysical simulations
such as core-collapse supernovae and binary neutron star mergers,
we construct a new EOS table (EOS5) using the TM1m model.
All quantities listed in the EOS table, as detailed in Appendix A of~\citet{shen11},
are calculated using the TM1m model for uniform matter at densities
higher than $\sim 10^{14}\,\mathrm{g/cm^{3}}$,
which can then be combined with the nonuniform part of the Shen EOS4 at low densities
to generate the final EOS table, as is often done in the literature~\citet{ishi08,sumi19}.
As illustrated in Figures~\ref{fig:21EA}-\ref{fig:23Esym}, the discrepancies
between the TM1m and TM1e models become increasingly evident as the density increases.
Consequently, our analysis focuses on a detailed comparison of uniform matter
at high densities. There will be a separate work on the low-density part of the EOS table elsewhere.

The EOS table covers a wide range of temperatures $T$, proton fractions $Y_p$,
and baryon mass densities $\rho_B$ for use in numerical simulations of
core-collapse supernovae, proto-neutron star cooling, and binary neutron star mergers.
For convenience in practical use, we provide the EOS table in the same
tabular form as given in Table 1 of~\citet{shen11}, which has been extensively
adopted by the EOSs posted on the public database CompOSE.
In general, when the temperature is higher than a critical value of $\sim 14$ MeV
and the density is beyond $\sim 10^{14.1}\,\rm{g/cm^{3}}$,
heavy nuclei dissolve and the favorable state is uniform nuclear matter.

\begin{figure}[htb]
\centering
\includegraphics[width=8.5 cm,clip]{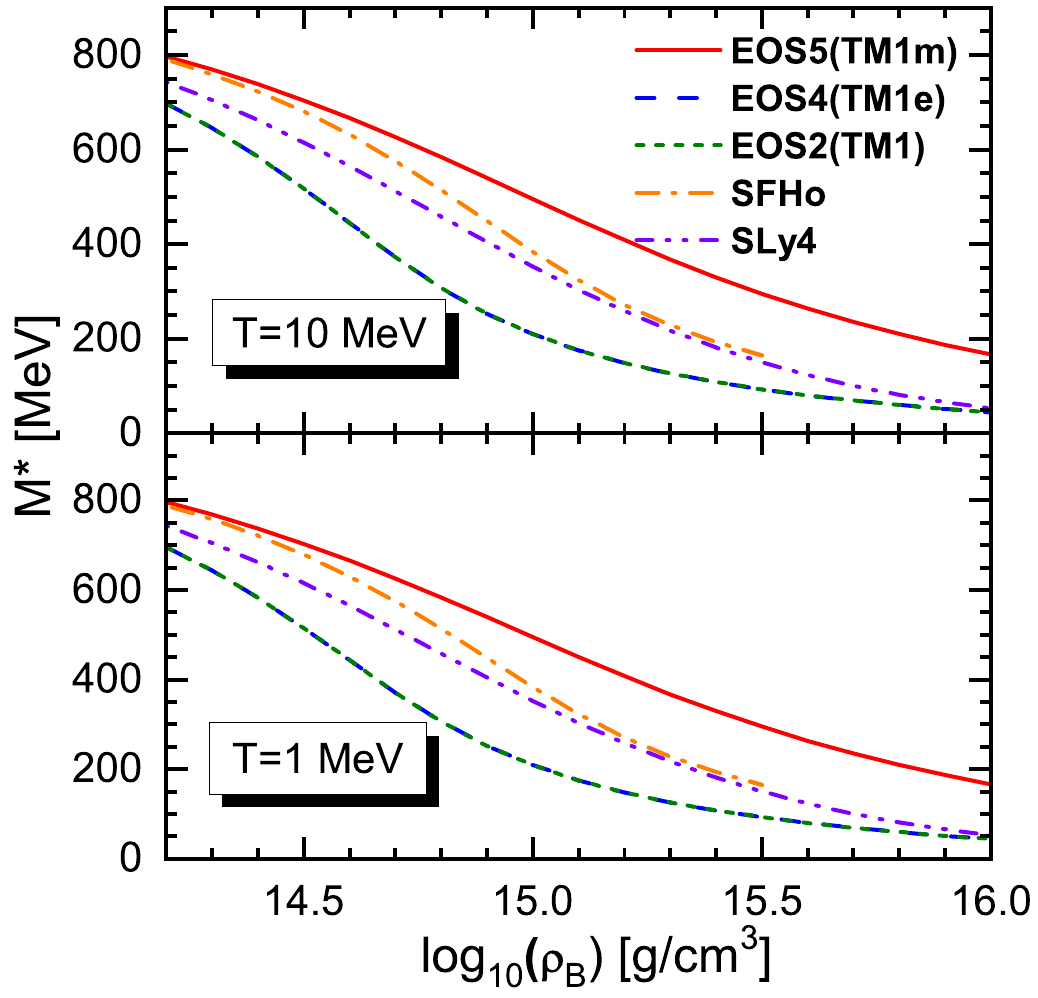}
\caption{Effective nucleon mass $M^{\ast}$ as a function of the baryon mass
density $\rho_B$ at $T=1$ and $10$ MeV.
The results are obtained for symmetric nuclear matter ($Y_p=0.5$).}

\label{fig:41MRho}
\end{figure}

In Figure~\ref{fig:41MRho}, we display the effective nucleon mass $M^{\ast}$
as a function of the baryon mass density $\rho_B$ at $T=1$ and $10$ MeV.
Although the calculations are performed at $Y_p=0.5$, the results of $M^{\ast}$
are found to be rather insensitive to $Y_p$. This is because $M^{\ast}$ is determined
by the scalar meson $\sigma$, according to the relation $M^{\ast}=M+g_{\sigma}\sigma$.
Notably, the effective masses in the TM1e and TM1 models are identical, due to
having the same isoscalar properties.
The results obtained in the TM1m model (red solid lines) are significantly larger
than those of TM1e (blue dashed lines) and TM1 (green dotted lines).
It is shown that $M^{\ast}$ decreases as the density increases.
At the saturation density $n_0$ ($\rho_B\simeq 10^{14.4}\,\rm{g/cm^{3}}$),
the effective masses obtained in the TM1m and TM1e models are, respectively,
$M^{\ast}=744$ MeV and $M^{\ast}=595$ MeV.
When the density rises to $\rho_B=10^{15}\,\rm{g/cm^{3}}$ (about $4n_0$),
the effective masses decrease to $M^{\ast}=496$ MeV in the TM1m model and
$M^{\ast}=210$ MeV in the TM1e model, respectively.

The effective masses obtained from the SFHo and SLy4 models show a similar
density dependence, while the values fall between those of the TM1m and TM1e models.
Comparing the case at $T=10$ MeV in the upper panel with that at $T=1$ MeV
in the lower panel, the behaviors of the effective masses are similar.
This suggests that the effective mass has a rather weak dependence on temperature.

\begin{figure}[htb]
\centering
\includegraphics[width=8.5 cm,clip]{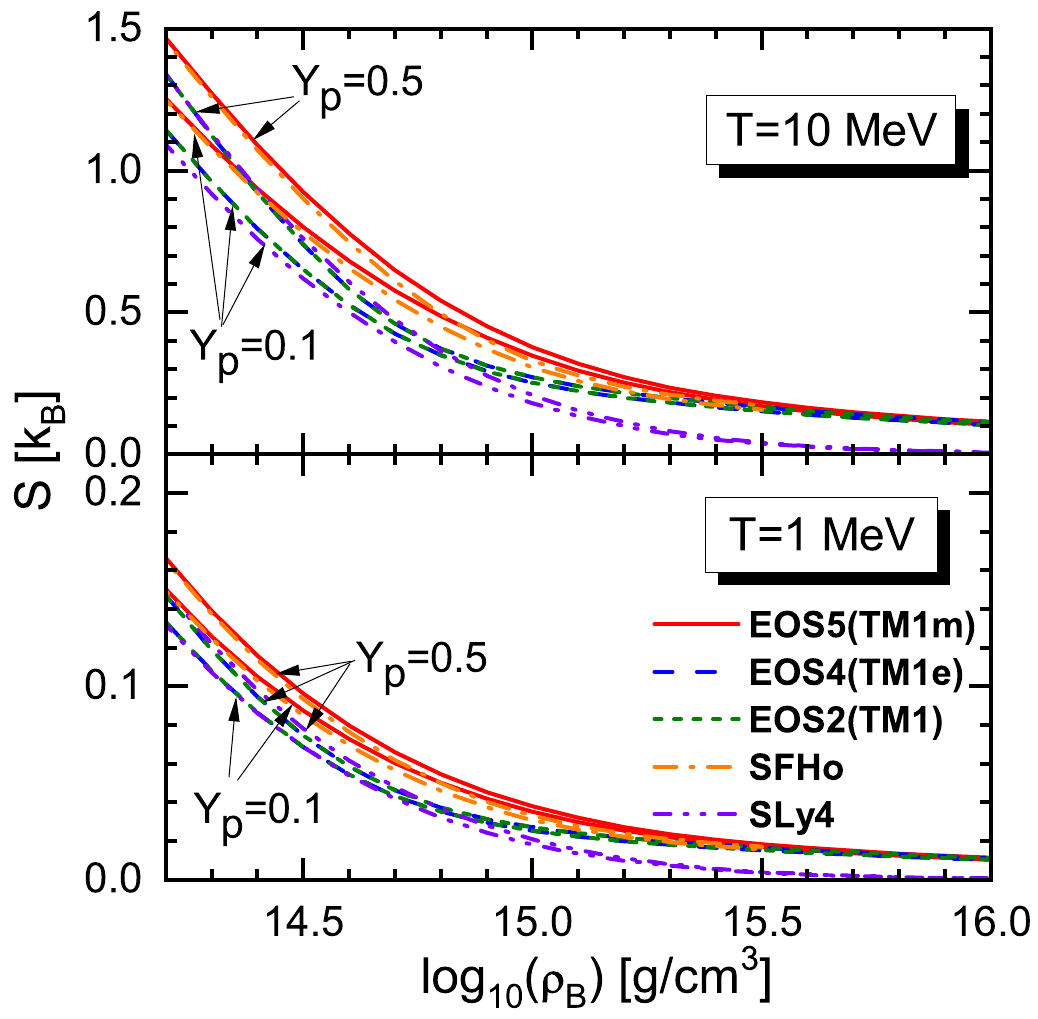}
\caption{Entropy per baryon $S$ as a function of the baryon mass
 density $\rho_B$ with $Y_p=0.1$ and $0.5$ at $T=1$ and $10$ MeV. }

\label{fig:42S}
\end{figure}
\begin{figure}[htb]
\centering
\includegraphics[width=8.5 cm,clip]{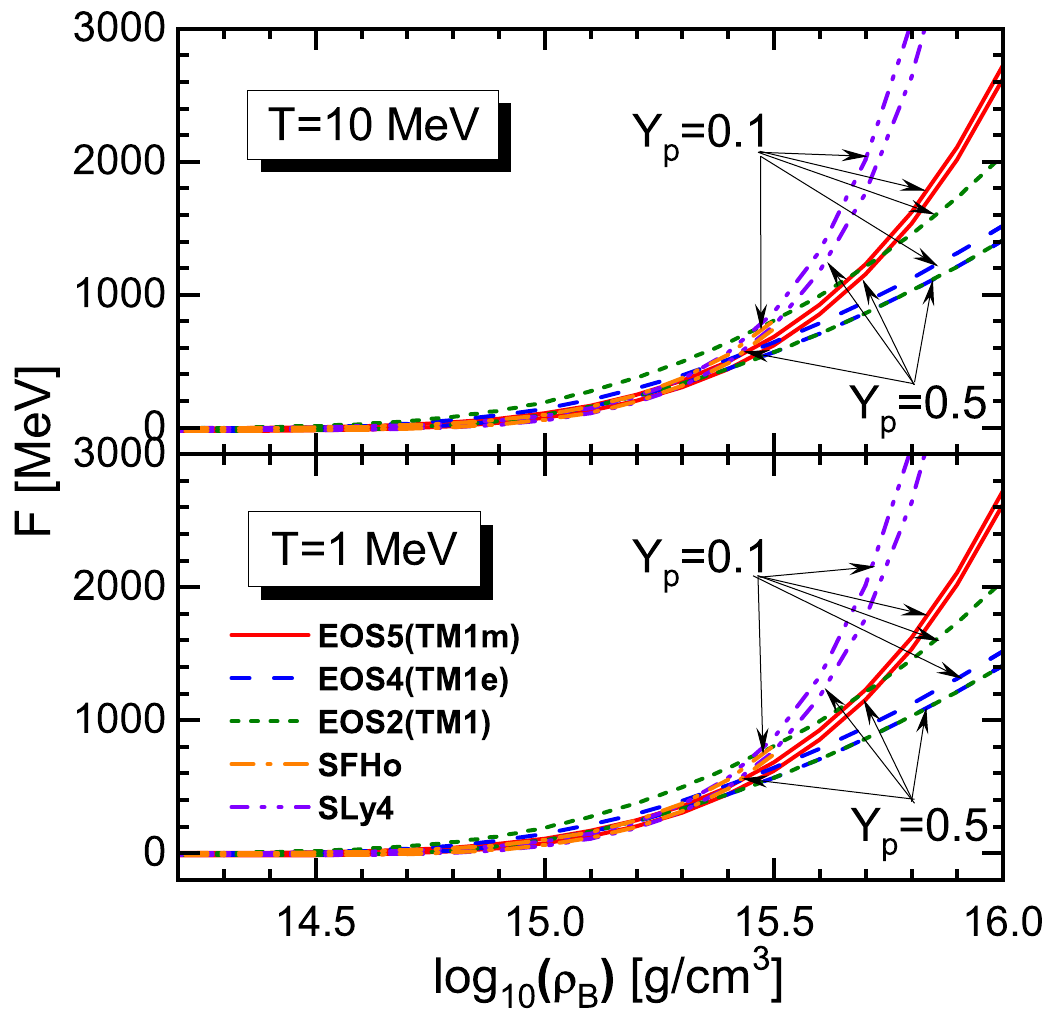}
\caption{The same as Figure~\ref{fig:42S}, but for the free energy per baryon $F$. }
\label{fig:43F}
\end{figure}
\begin{figure}[htb]
\centering
\includegraphics[width=8.5 cm,clip]{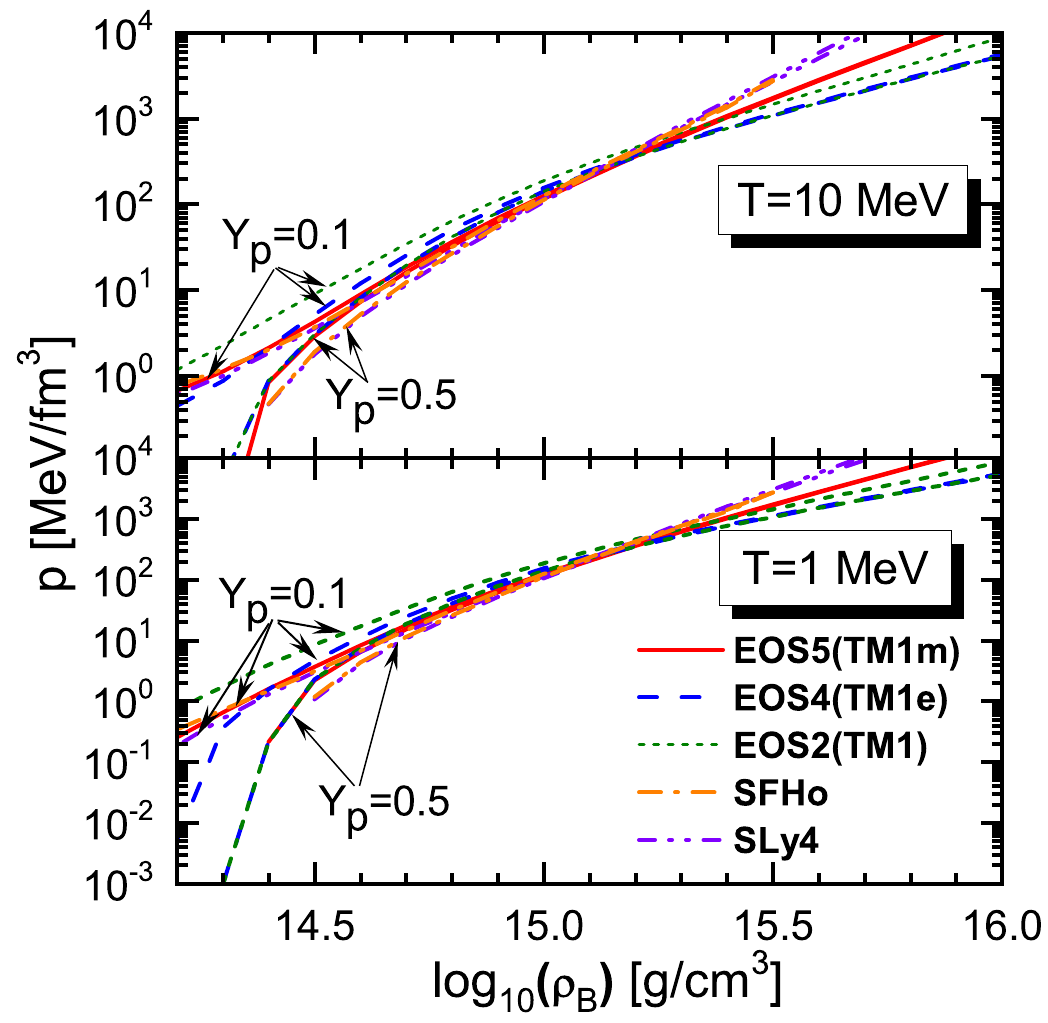}
\caption{The same as Figure~\ref{fig:42S}, but for the pressure $p$.    \,\,}
\label{fig:44P}
\end{figure}

It is essential to investigate how the effective mass influences the thermodynamic quantities within the EOS table.

In Figure~\ref{fig:42S}, we show the entropy per baryon $S$ as a function of
$\rho_B$ for $Y_p=0.1$ and $0.5$ at $T=1$ and $10$ MeV. It is observed that
the values of $S$ significantly decrease with increasing density.
One can see that, around the saturation density, there are noticeable differences
among the models considered in this work. However, the differences among all RMF
models (TM1, TM1e, TM1m, and SFHo) nearly disappear at higher densities.
Meanwhile, the deviation between the nonrelativistic SLy4 model and the RMF models
remains finite, especially in the high-density region.
This deviation is caused by a nonrelativistic calculation for $S$ used
in the SLy4 model~\citep{schn17}.
When the relativistic formula $\sqrt{k^2+{M^{\ast}}^2}$ in Equation~(\ref{eq:firmf})
is simply replaced by a nonrelativistic approximation $M^{\ast}+k^2/2M^{\ast}$,
the entropy per baryon $S$ obtained in the TM1m model gradually approaches
the results of the nonrelativistic SLy4 model with increasing density,
which verifies that the deviation in $S$ at higher densities between the TM1m
and SLy4 models is mainly due to relativistic effects.
From Equations~(\ref{eq:SRMF}) and~(\ref{eq:firmf}), it is seen that
the entropy $S$ is closely related to the effective mass $M^{\ast}$.
The discrepancies between these models can be understood by analyzing the
behaviors of $M^{\ast}$, as shown in Figure~\ref{fig:41MRho}.
On the other hand, the difference of the  symmetry energy between the TM1 and TM1e
models has a minor influence on the entropy.
It is worth noting that the entropy values at $T=10$ MeV (upper panel)
are roughly ten times greater than those of $T=1$ MeV (lower panel).
Generally, the differences between $Y_p=0.1$ and $0.5$ decrease
as the density increases.

In Figure~\ref{fig:43F}, we plot the free energy per baryon $F$ as a function of
the baryon mass density $\rho_B$ for $Y_p=0.1$ and $0.5$ at $T=1$ and $10$ MeV.
We first compare the results in EOS5 (TM1m) with those in EOS4 (TM1e) and EOS2 (TM1).
At densities below $\rho_B \simeq 10^{15.4}\,\rm{g/cm^{3}}$ (about $10n_0$),
the free energies in the TM1m model are lower than those in the TM1e model,
which can be more easily observed in Figure~\ref{fig:21EA}.
This is because the larger effective mass in the TM1m model leads to a relatively
smaller kinetic energy. However, at extremely high densities, the free energies
in the TM1m model exceed those in the TM1e model, owing to the growing
contributions from nonlinear meson terms.

For comparison, the results obtained from the SFHo and SLy4 models are also
plotted in this figure. At densities below $\rho_B \simeq 10^{15.5}\,\rm{g/cm^{3}}$,
the behaviors of $F$ in the SFHo and SLy4 models are similar to those in the TM1m
model, due to their similar saturation properties.
When $\rho_B > 10^{15.5}\,\rm{g/cm^{3}}$, the SFHo model fails to obtain
physical solutions for the field equations. Consequently, the maximum density
in the SFHo EOS table on CompOSE is set at $10^{15.5}\,\rm{g/cm^{3}}$ ($1.9$ fm$^{-3}$).
On the other hand, the nonrelativistic SLy4 model yields significantly larger
$F$ than the TM1m model at densities higher than $\sim 10^{15.5}\,\rm{g/cm^{3}}$.
This discrepancy is primarily due to the distinct density-dependent behaviors,
which gradually become more apparent as the density increases.
As for the dependence on $Y_p$, it is noteworthy that the differences between
the results of $Y_p=0.5$ and $Y_p=0.1$ are clearly model-dependent.
In contrast to other models, the original TM1 model presents considerably
larger differences between $Y_p=0.5$ and $Y_p=0.1$. This is because
the TM1 model exhibits significantly larger symmetry energies at higher
densities, as shown in Figure~\ref{fig:23Esym}.
Comparing the free-energy trends between $T=1$ MeV and $T=10$ MeV,
it is observed that they are very similar.

In Figure~\ref{fig:44P}, we display the pressure $p$ as a function of $\rho_B$
for $Y_p=0.1$ and $0.5$ at $T=1$ and $10$ MeV.
According to tradition, when constructing an EOS table,
the contributions from leptons and photons are usually calculated separately,
so they are not included in the present work. As shown in Figure~\ref{fig:44P},
it is evident that at low densities, the pressure values are very small,
even reaching negative levels when $Y_p=0.5$.
It is known that the pressure of leptons and photons is significantly higher than
that of baryons around the saturation density, and the total pressure
becomes positive when the contributions from leptons and photons are taken into account.
With increasing density, the pressure rises rapidly.

The discrepancies between the different models are density-dependent.
At densities below $\rho_B \simeq 10^{15.2}\,\rm{g/cm^{3}}$,
the pressures in the TM1 model with $Y_p=0.1$ are visibly higher than those
in other models, which is caused by the larger symmetry energy slope $L$.
It has been pointed out in~\citet{oyam07} that a larger value of $L$ implies
a higher pressure of pure neutron matter.
At extremely high densities, the pressure values in the different models exhibit
distinct density-dependent behaviors.
This is due to the increasing influence of nonlinear meson terms.

\section{Summary}
\label{sec:4}

In this work, we have investigated the influence of the effective nucleon mass on
the EOS for astrophysical simulations such as core-collapse supernovae,
proto-neutron star cooling, and binary neutron star mergers.
We have employed the RMF model, in which the effective
nucleon mass is more consistently integrated than it is in nonrelativistic
approaches such as the Skyrme model.
A new RMF parameter set, TM1m, is introduced in the present work, which is
a modification of the TM1e parameterization with a larger effective mass.
The TM1m model preserves the same saturation properties as the TM1e model,
but their effective masses are significantly different:
$M^{\ast}/M \sim 0.8$ for the TM1m model and $M^{\ast}/M \sim 0.63$ for the TM1e model.
It is well known that there exists a positive correlation between the
symmetry energy slope $L$ and the neutron star radius.
In contrast to the original TM1 model characterized by $L=111$ MeV,
both the TM1m and TM1e models provide a small value of $L=40$ MeV,
which is more favored by recent astrophysical observations.

We calculate the properties of cold neutron stars using the TM1m model
and compare with the results obtained by the TM1e and TM1 models.

For comparison, we also include the results from an RMF model with the SFHo
parameterization and a nonrelativistic Skyrme model with the SLy4 parameterization.
The maximum masses of neutron stars obtained in these models could fulfill
the observational constraint $M_{\mathrm{max}} > 2\ M_\odot$.
It is found that the mass-radius relations exhibit significant variations
among these models, which might be caused by different behaviors of the 
symmetry energy and its slope.
Regarding the radius of a canonical $1.4\ M_\odot$ neutron star ($R_{1.4}$),
a large value of $\sim 14.2$ km is obtained using the TM1 model, which is 
reduced to $13.1$ km in the TM1e model and $12.4$ km in the TM1m models, respectively.
The resulting radius $R_{1.4}$ and the tidal deformability $\Lambda_{1.4}$
within the TM1m model could be compatible with the constraint
inferred from the gravitational-wave event GW170817.

To explore the impact of the effective nucleon mass in astrophysical simulations
such as core-collapse supernovae and binary neutron star mergers,
we have constructed a new EOS table (EOS5) using the TM1m model.
All quantities included in the EOS table are calculated using the TM1m model
for uniform matter at densities higher than $\sim 10^{14}\,\mathrm{g/cm^{3}}$,
which could be combined with the nonuniform part of the Shen EOS4 at low densities
to generate the final EOS table.
It is shown that the effective nucleon mass $M^{\ast}$ significantly decreases
as the density increases.
At supra-saturation densities, the TM1m model predicts lower free energies and
pressures than the TM1e model, which may be due to the fact that the larger effective
masses in the TM1m model result in relatively smaller kinetic energies.
However, at extremely high densities, the free energies in the TM1m model
exceed those in the TM1e model, owing to the growing contributions from
nonlinear meson terms.

Compared to the TM1m model, the SFHo model has similar saturation properties,
with $M^{\ast}/M \sim 0.76$, which yield comparable results in the EOS table.
At higher densities, the nonrelativistic SLy4 model predicts smaller entropy
and larger free energy than the RMF models.
The new EOS table (EOS5), constructed using the TM1m model, offers a valuable
resource for exploring the influence of the  effective nucleon mass on the dynamics
of supernovae within the same parameterization family.
The advantage of conducting a comparison between the TM1m and TM1e EOS
tables lies in the fact that only the effective masses differ while the other
saturation properties remain identical.
Numerical simulations of core-collapse supernovae, as well as analyses of
the influence of the effective nucleon mass, are currently underway.

\section{Acknowledgments}

This work was supported in part by the National Natural Science Foundation of
China (grant Nos. 12175109 and 12475149) and the Natural Science Foundation
of Guangdong Province (grant No: 2024A1515010911).
K.S. is supported by Grant-in-Aid for Scientific Research (19K03837, 20H01905, and 24K00632)
from the Ministry of Education, Culture, Sports, Science and Technology (MEXT), Japan.
K.S. acknowledges the Computing Research Center, KEK, JLDG on SINET of NII,
the Research Center for Nuclear Physics, Osaka University, the Yukawa Institute of
Theoretical Physics, Kyoto University, Nagoya University, and the Information
Technology Center, University of Tokyo, for providing high-performance computing resources.
K.S. is partly supported by MEXT via the 
\textquotedblleft Program for Promoting Researches on the Supercomputer Fugaku\textquotedblright\ (Structure and Evolution of the Universe Unraveled
by the Fusion of Simulation and AI, JPMXP1020230406,
Project IDs: hp230204, hp230270, hp240219, and hp240264),
the HPCI System Research Project (Project IDs: hp230056 and hp240041),
and the Particle, Nuclear
and Astro Physics Simulation Program (No. 2023-003) of the Institute of Particle
and Nuclear Studies, High Energy Accelerator Research Organization (KEK).

\bibliographystyle{aasjournal}
	
\bibliography{refs-shen}

\begin{thebibliography}{}
\expandafter\ifx\csname natexlab\endcsname\relax\def\natexlab#1{#1}\fi
\providecommand{\url}[1]{\href{#1}{#1}}
\providecommand{\dodoi}[1]{doi:~\href{http://doi.org/#1}{\nolinkurl{#1}}}
\providecommand{\doeprint}[1]{\href{http://ascl.net/#1}{\nolinkurl{http://ascl.net/#1}}}
\providecommand{\doarXiv}[1]{\href{https://arxiv.org/abs/#1}{\nolinkurl{https://arxiv.org/abs/#1}}}

\bibitem[{Abbott {et~al.}(2017)Abbott, Abbott, Abbott, Acernese, Ackley, Adams,
  Adams, Addesso, Adhikari, Adya, {et~al.}}]{abbo17}
Abbott, B.~P., Abbott, R., Abbott, T., {et~al.} 2017, Physical Review Letters,
  119, 161101

\bibitem[{Abbott {et~al.}(2018)Abbott, Abbott, Abbott, Acernese, Ackley, Adams,
  Adams, Addesso, Adhikari, Adya, {et~al.}}]{abbo18}
---. 2018, Physical Review Letters, 121, 161101,
  \dodoi{10.1103/PhysRevLett.121.161101}

\bibitem[{Andersen {et~al.}(2021)Andersen, Zha, da~Silva~Schneider,
  Betranhandy, Couch, \& O¡¯Connor}]{ande21}
Andersen, O.~E., Zha, S., da~Silva~Schneider, A., {et~al.} 2021, The
  Astrophysical Journal, 923, 201, \dodoi{10.3847/1538-4357/ac294c}

\bibitem[{Antoniadis {et~al.}(2013)Antoniadis, Freire, Wex, Tauris, Lynch, van
  Kerkwijk, Kramer, Bassa, Dhillon, Driebe, Hessels, Kaspi, Kondratiev, Langer,
  Marsh, McLaughlin, Pennucci, Ransom, Stairs, van Leeuwen, Verbiest, \&
  Whelan}]{anto13}
Antoniadis, J., Freire, P. C.~C., Wex, N., {et~al.} 2013, Science, 340,
  1233232, \dodoi{10.1126/science.1233232}

\bibitem[{Arzoumanian {et~al.}(2018)Arzoumanian, Brazier, Burke-Spolaor,
  Chamberlin, Chatterjee, Christy, Cordes, Cornish, Crawford, Cromartie,
  Crowter, DeCesar, Demorest, Dolch, Ellis, Ferdman, Ferrara, Fonseca,
  Garver-Daniels, Gentile, Halmrast, Huerta, Jenet, Jessup, Jones, Jones,
  Kaplan, Lam, Lazio, Levin, Lommen, Lorimer, Luo, Lynch, Madison, Matthews,
  McLaughlin, McWilliams, Mingarelli, Ng, Nice, Pennucci, Ransom, Ray, Siemens,
  Simon, Spiewak, Stairs, Stinebring, Stovall, Swiggum, Taylor, Vallisneri, van
  Haasteren, Vigeland, Zhu, \& Collaboration}]{arzo18}
Arzoumanian, Z., Brazier, A., Burke-Spolaor, S., {et~al.} 2018, The
  Astrophysical Journal Supplement Series, 235, 37,
  \dodoi{10.3847/1538-4365/aab5b0}

\bibitem[{Bao {et~al.}(2014)Bao, Hu, Zhang, \& Shen}]{bao14b}
Bao, S., Hu, J., Zhang, Z., \& Shen, H. 2014, Physical Review C, 90, 045802,
  \dodoi{10.1103/PhysRevC.90.045802}

\bibitem[{Baym {et~al.}(1971)Baym, Bethe, \& Pethick}]{baym71}
Baym, G., Bethe, H.~A., \& Pethick, C.~J. 1971, Nuclear Physics A, 175, 225,
  \dodoi{https://doi.org/10.1016/0375-9474(71)90281-8}

\bibitem[{Chabanat {et~al.}(1998)Chabanat, Bonche, Haensel, Meyer, \&
  Schaeffer}]{chab98}
Chabanat, E., Bonche, P., Haensel, P., Meyer, J., \& Schaeffer, R. 1998,
  Nuclear Physics A, 635, 231,
  \dodoi{https://doi.org/10.1016/S0375-9474(98)00180-8}

\bibitem[{Chen \& Piekarewicz(2014)}]{chen14}
Chen, W.-C., \& Piekarewicz, J. 2014, Physical Review C, 90, 044305,
  \dodoi{10.1103/PhysRevC.90.044305}

\bibitem[{Chen \& Piekarewicz(2015)}]{chen15}
---. 2015, Physics Letters B, 748, 284,
  \dodoi{https://doi.org/10.1016/j.physletb.2015.07.020}

\bibitem[{Choi {et~al.}(2021)Choi, Miyatsu, Cheoun, \& Saito}]{choi21}
Choi, S., Miyatsu, T., Cheoun, M.-K., \& Saito, K. 2021, The Astrophysical
  Journal, 909, 156, \dodoi{10.3847/1538-4357/abe3fe}

\bibitem[{Dutra {et~al.}(2014)Dutra, Louren{\c{c}}o, Avancini, Carlson,
  Delfino, Menezes, Provid{\^e}ncia, Typel, \& Stone}]{dutr14}
Dutra, M., Louren{\c{c}}o, O., Avancini, S., {et~al.} 2014, Physical Review C,
  90, 055203

\bibitem[{Fantina {et~al.}(2013)Fantina, Chamel, Pearson, \& Goriely}]{fant13}
Fantina, A., Chamel, N., Pearson, J., \& Goriely, S. 2013, Astronomy \&
  astrophysics, 559, A128, \dodoi{10.1051/0004-6361/201321884}

\bibitem[{Fattoyev {et~al.}(2010)Fattoyev, Horowitz, Piekarewicz, \&
  Shen}]{fatt10}
Fattoyev, F.~J., Horowitz, C.~J., Piekarewicz, J., \& Shen, G. 2010, Physical
  Review C, 82, 055803, \dodoi{10.1103/PhysRevC.82.055803}

\bibitem[{Fattoyev {et~al.}(2018)Fattoyev, Piekarewicz, \& Horowitz}]{fatt18}
Fattoyev, F.~J., Piekarewicz, J., \& Horowitz, C.~J. 2018, Physical Review
  Letters, 120, 172702, \dodoi{10.1103/PhysRevLett.120.172702}

\bibitem[{Fonseca {et~al.}(2021)Fonseca, Cromartie, Pennucci, Ray, Kirichenko,
  Ransom, Demorest, Stairs, Arzoumanian, Guillemot, Parthasarathy, Kerr,
  Cognard, Baker, Blumer, Brook, DeCesar, Dolch, Dong, Ferrara, Fiore,
  Garver-Daniels, Good, Jennings, Jones, Kaspi, Lam, Lorimer, Luo, McEwen,
  McKee, McLaughlin, McMann, Meyers, Naidu, Ng, Nice, Pol, Radovan,
  Shapiro-Albert, Tan, Tendulkar, Swiggum, Wahl, \& Zhu}]{fons21}
Fonseca, E., Cromartie, H.~T., Pennucci, T.~T., {et~al.} 2021, The
  Astrophysical Journal Letters, 915, L12, \dodoi{10.3847/2041-8213/ac03b8}

\bibitem[{Fortin {et~al.}(2021)Fortin, Raduta, Avancini, \&
  Provid\^encia}]{fort21}
Fortin, M., Raduta, A.~R., Avancini, S., \& Provid\^encia, C. m.~c. 2021,
  Physical Review D, 103, 083004, \dodoi{10.1103/PhysRevD.103.083004}

\bibitem[{Furusawa {et~al.}(2017{\natexlab{a}})Furusawa, Sumiyoshi, Yamada, \&
  Suzuki}]{furu17a}
Furusawa, S., Sumiyoshi, K., Yamada, S., \& Suzuki, H. 2017{\natexlab{a}},
  Nuclear Physics A, 957, 188,
  \dodoi{https://doi.org/10.1016/j.nuclphysa.2016.09.002}

\bibitem[{Furusawa {et~al.}(2017{\natexlab{b}})Furusawa, Togashi, Nagakura,
  Sumiyoshi, Yamada, Suzuki, \& Takano}]{furu17b}
Furusawa, S., Togashi, H., Nagakura, H., {et~al.} 2017{\natexlab{b}}, Journal
  of Physics G: Nuclear and Particle Physics, 44, 094001

\bibitem[{Furusawa {et~al.}(2020)Furusawa, Togashi, Sumiyoshi, Saito, Yamada,
  \& Suzuki}]{furu20}
Furusawa, S., Togashi, H., Sumiyoshi, K., {et~al.} 2020, Progress of
  Theoretical and Experimental Physics, 2020, 013D05,
  \dodoi{10.1093/ptep/ptz135}

\bibitem[{Hempel \& Schaffner-Bielich(2010)}]{hemp10}
Hempel, M., \& Schaffner-Bielich, J. 2010, Nuclear Physics A, 837, 210,
  \dodoi{https://doi.org/10.1016/j.nuclphysa.2010.02.010}

\bibitem[{Huang {et~al.}(2022{\natexlab{a}})Huang, Hu, Zhang, \&
  Shen}]{huang22a}
Huang, K., Hu, J., Zhang, Y., \& Shen, H. 2022{\natexlab{a}}, The Astrophysical
  Journal, 935, 88, \dodoi{10.3847/1538-4357/ac7f3c}

\bibitem[{Huang {et~al.}(2022{\natexlab{b}})Huang, Hu, Zhang, \&
  Shen}]{huang22b}
---. 2022{\natexlab{b}}, Nuclear Physics Review, 39, 135,
  \dodoi{10.11804/NuclPhysRev.39.2022013}

\bibitem[{Ishizuka {et~al.}(2008)Ishizuka, Ohnishi, Tsubakihara, Sumiyoshi, \&
  Yamada}]{ishi08}
Ishizuka, C., Ohnishi, A., Tsubakihara, K., Sumiyoshi, K., \& Yamada, S. 2008,
  Journal of Physics G: Nuclear and Particle Physics, 35, 085201,
  \dodoi{10.1088/0954-3899/35/8/085201}

\bibitem[{Ji {et~al.}(2019)Ji, Hu, Bao, \& Shen}]{ji19}
Ji, F., Hu, J., Bao, S., \& Shen, H. 2019, Physical Review C, 100, 045801

\bibitem[{Lalazissis {et~al.}(1997)Lalazissis, K{\"o}nig, \& Ring}]{lala97}
Lalazissis, G., K{\"o}nig, J., \& Ring, P. 1997, Physical Review C, 55, 540

\bibitem[{Lattimer {et~al.}(1991)Lattimer, Pethick, Prakash, \&
  Haensel}]{latt91b}
Lattimer, J.~M., Pethick, C.~J., Prakash, M., \& Haensel, P. 1991, Physical
  Review Letters, 66, 2701, \dodoi{10.1103/PhysRevLett.66.2701}

\bibitem[{Lattimer \& Swesty(1991)}]{latt91}
Lattimer, J.~M., \& Swesty, F.~D. 1991, Nuclear Physics A, 535, 331,
  \dodoi{https://doi.org/10.1016/0375-9474(91)90452-C}

\bibitem[{Lim {et~al.}(2017)Lim, Hyun, \& Lee}]{lim17}
Lim, Y., Hyun, C.~H., \& Lee, C.-H. 2017, International Journal of Modern
  Physics E, 26, 1750015

\bibitem[{Maruyama {et~al.}(2007)Maruyama, Chiba, Schulze, \& Tatsumi}]{maru07}
Maruyama, T., Chiba, S., Schulze, H.-J., \& Tatsumi, T. 2007, Physical Review
  D, 76, 123015, \dodoi{10.1103/PhysRevD.76.123015}

\bibitem[{Miller {et~al.}(2019)Miller, Lamb, Dittmann, Bogdanov, Arzoumanian,
  Gendreau, Guillot, Harding, Ho, Lattimer, {et~al.}}]{mill19}
Miller, M.~C., Lamb, F.~K., Dittmann, A.~J., {et~al.} 2019, The Astrophysical
  Journal Letters, 887, L24, \dodoi{10.3847/2041-8213/ab50c5}

\bibitem[{Miller {et~al.}(2021)Miller, Lamb, Dittmann, Bogdanov, Arzoumanian,
  Gendreau, Guillot, Ho, Lattimer, Loewenstein, {et~al.}}]{mill21}
---. 2021, The Astrophysical Journal Letters, 918, L28,
  \dodoi{10.3847/2041-8213/ac089b}

\bibitem[{Miyatsu {et~al.}(2024)Miyatsu, Cheoun, Kim, \& Saito}]{miya24}
Miyatsu, T., Cheoun, M.-K., Kim, K., \& Saito, K. 2024, Novel features of
  asymmetric nuclear matter from large neutron skin thickness and small
  neutron-star radii.
\newblock \doarXiv{2411.13210}

\bibitem[{Most {et~al.}(2018)Most, Weih, Rezzolla, \&
  Schaffner-Bielich}]{most18}
Most, E.~R., Weih, L.~R., Rezzolla, L., \& Schaffner-Bielich, J. 2018, Physical
  Review Letters, 120, 261103, \dodoi{10.1103/PhysRevLett.120.261103}

\bibitem[{Nagakura {et~al.}(2019)Nagakura, Furusawa, Togashi, Richers,
  Sumiyoshi, \& Yamada}]{naga19}
Nagakura, H., Furusawa, S., Togashi, H., {et~al.} 2019, The Astrophysical
  Journal Supplement Series, 240, 38, \dodoi{10.3847/1538-4365/aafac9}

\bibitem[{Nakazato \& Suzuki(2019)}]{naka19}
Nakazato, K., \& Suzuki, H. 2019, The Astrophysical Journal, 878, 25,
  \dodoi{10.3847/1538-4357/ab1d4b}

\bibitem[{Oertel {et~al.}(2017)Oertel, Hempel, Kl\"ahn, \& Typel}]{oert17}
Oertel, M., Hempel, M., Kl\"ahn, T., \& Typel, S. 2017, Reviews of Modern
  Physics, 89, 015007, \dodoi{10.1103/RevModPhys.89.015007}

\bibitem[{Oyamatsu \& Iida(2007)}]{oyam07}
Oyamatsu, K., \& Iida, K. 2007, Physical Review C, 75, 015801,
  \dodoi{10.1103/PhysRevC.75.015801}

\bibitem[{Raduta \& Gulminelli(2019)}]{radu19}
Raduta, A., \& Gulminelli, F. 2019, Nuclear Physics A, 983, 252,
  \dodoi{https://doi.org/10.1016/j.nuclphysa.2018.11.003}

\bibitem[{Riley {et~al.}(2019)Riley, Watts, Bogdanov, Ray, Ludlam, Guillot,
  Arzoumanian, Baker, Bilous, Chakrabarty, {et~al.}}]{rile19}
Riley, T.~E., Watts, A.~L., Bogdanov, S., {et~al.} 2019, The Astrophysical
  Journal Letters, 887, L21

\bibitem[{Riley {et~al.}(2021)Riley, Watts, Ray, Bogdanov, Guillot, Morsink,
  Bilous, Arzoumanian, Choudhury, Deneva, {et~al.}}]{rile21}
Riley, T.~E., Watts, A.~L., Ray, P.~S., {et~al.} 2021, The Astrophysical
  Journal Letters, 918, L27

\bibitem[{Schneider {et~al.}(2017)Schneider, Roberts, \& Ott}]{schn17}
Schneider, A.~S., Roberts, L.~F., \& Ott, C.~D. 2017, Physical Review C, 96,
  065802, \dodoi{10.1103/PhysRevC.96.065802}

\bibitem[{Schneider {et~al.}(2019)Schneider, Roberts, Ott, \&
  O'Connor}]{schn19}
Schneider, A.~S., Roberts, L.~F., Ott, C.~D., \& O'Connor, E. 2019, Physical
  Review C, 100, 055802, \dodoi{10.1103/PhysRevC.100.055802}

\bibitem[{Serot \& Walecka(1986)}]{sero86}
Serot, B.~D., \& Walecka, J.~D. 1986, Advances in Nuclear Physics, edited by JW
  Negele and E. Vogt,  Plenum, New York

\bibitem[{Shen {et~al.}(2011{\natexlab{a}})Shen, Horowitz, \&
  O'Connor}]{shenG11a}
Shen, G., Horowitz, C.~J., \& O'Connor, E. 2011{\natexlab{a}}, Physical Review
  C, 83, 065808, \dodoi{10.1103/PhysRevC.83.065808}

\bibitem[{Shen {et~al.}(2020)Shen, Ji, Hu, \& Sumiyoshi}]{shen20}
Shen, H., Ji, F., Hu, J., \& Sumiyoshi, K. 2020, The Astrophysical Journal,
  891, 148, \dodoi{10.3847/1538-4357/ab72fd}

\bibitem[{Shen {et~al.}(1998{\natexlab{a}})Shen, Toki, Oyamatsu, \&
  Sumiyoshi}]{shen98b}
Shen, H., Toki, H., Oyamatsu, K., \& Sumiyoshi, K. 1998{\natexlab{a}}, Progress
  of Theoretical Physics, 100, 1013

\bibitem[{Shen {et~al.}(1998{\natexlab{b}})Shen, Toki, Oyamatsu, \&
  Sumiyoshi}]{shen98a}
---. 1998{\natexlab{b}}, Nuclear Physics A, 637, 435

\bibitem[{Shen {et~al.}(2011{\natexlab{b}})Shen, Toki, Oyamatsu, \&
  Sumiyoshi}]{shen11}
---. 2011{\natexlab{b}}, The Astrophysical Journal Supplement Series, 197, 20,
  \dodoi{10.1088/0067-0049/197/2/20}

\bibitem[{Steiner {et~al.}(2013)Steiner, Hempel, \& Fischer}]{stei13}
Steiner, A.~W., Hempel, M., \& Fischer, T. 2013, The Astrophysical Journal,
  774, 17, \dodoi{10.1088/0004-637X/774/1/17}

\bibitem[{Sumiyoshi {et~al.}(2023)Sumiyoshi, Kojo, \& Furusawa}]{sumi23}
Sumiyoshi, K., Kojo, T., \& Furusawa, S. 2023, Equation of State in Neutron
  Stars and Supernovae (Springer Nature Singapore), 1--51,
  \dodoi{10.1007/978-981-15-8818-1_104-1}

\bibitem[{Sumiyoshi {et~al.}(2019)Sumiyoshi, Nakazato, Suzuki, Hu, \&
  Shen}]{sumi19}
Sumiyoshi, K., Nakazato, K., Suzuki, H., Hu, J., \& Shen, H. 2019, The
  Astrophysical Journal, 887, 110, \dodoi{10.3847/1538-4357/ab5443}

\bibitem[{Togashi {et~al.}(2017)Togashi, Nakazato, Takehara, Yamamuro, Suzuki,
  \& Takano}]{toga17}
Togashi, H., Nakazato, K., Takehara, Y., {et~al.} 2017, Nuclear Physics A, 961,
  78.
\newblock \url{https://dx.doi.org/10.1016/j.nuclphysa.2017.02.010}

\bibitem[{Typel \& Alvear~Terrero(2020)}]{type20}
Typel, S., \& Alvear~Terrero, D. 2020, The European Physical Journal A, 56,
  160, \dodoi{10.1140/epja/s10050-020-00172-2}

\bibitem[{Typel {et~al.}(2015)Typel, Oertel, \& Kl{\"a}hn}]{type15}
Typel, S., Oertel, M., \& Kl{\"a}hn, T. 2015, Physics of Particles and Nuclei,
  46, 633

\bibitem[{Typel {et~al.}(2022)Typel, Oertel, Kl{\"a}hn, Chatterjee, Dexheimer,
  Ishizuka, Mancini, Novak, Pais, {et~al.}}]{type22}
Typel, S., Oertel, M., Kl{\"a}hn, T., {et~al.} 2022, The European Physical
  Journal A, 58, 221.
\newblock \url{https://compose.obspm.fr/}

\bibitem[{Weber {et~al.}(2019)Weber, Farrell, Spinella, Malfatti, Orsaria,
  Contrera, \& Maloney}]{webe19}
Weber, F., Farrell, D., Spinella, W.~M., {et~al.} 2019, Universe, 5, 169.
\newblock \url{https://dx.doi.org/10.3390/universe5070169}

\bibitem[{Yasin {et~al.}(2020)Yasin, Sch\"afer, Arcones, \& Schwenk}]{yasi20}
Yasin, H., Sch\"afer, S., Arcones, A., \& Schwenk, A. 2020, Physical Review
  Letters, 124, 092701, \dodoi{10.1103/PhysRevLett.124.092701}

\bibitem[{Yasutake {et~al.}(2014)Yasutake, {\L}astowiecki, Beni{\'c}, Blaschke,
  Maruyama, \& Tatsumi}]{yasu14}
Yasutake, N., {\L}astowiecki, R., Beni{\'c}, S., {et~al.} 2014, Physical Review
  C, 89, 065803, \dodoi{10.1103/PhysRevC.89.065803}

\bibitem[{Zhang \& Chen(2013)}]{zhang13}
Zhang, Z., \& Chen, L.-W. 2013, Physics Letters B, 726, 234,
  \dodoi{https://doi.org/10.1016/j.physletb.2013.08.002}

\end{thebibliography}

\end{document}